# Electrostatically Controlled Pyrophototronic Effect Enabled Accident Alert System using a Strain-Polarized WS$_2$ Phototransistor

Poulomi Chakrabarty,[1] Sera Sen,[1] Anwesha Chakraborty,[2] Abhay Anand VS,[3] Srilagna Sahoo,[1] Anshuman Kumar,[3] Debjani Karmakar,[2,4,5] and Saurabh Lodha[1*]

[1]Department of Electrical Engineering, Indian Institute of Technology Bombay, India
, [2]Technical Physics Division, BARC, Mumbai 400085, India, [3]Department of Physics, Indian Institute of Technology Bombay, India, Homi Bhabha National Institute, [4]Anushaktinagar, Mumbai, 400094, India, Department of Physics and Astronomy, [5]Uppsala University, Box 516, SE-75120, Uppsala, Sweden
Email: slodha@ee.iitb.ac.in

**Abstract**

Event-based dynamic light detection, specifically in low illumination power environments, is a critical requirement in autonomous vehicles. This work reports low optical power photodetection through the dynamic pyrophototronic effect in an ultra-thin 2D WS$_2$ phototransistor. A four-stage pyrophototronic photoresponse has been realized through biaxial strain-polarization of the non-centrosymmetric (5-layer) WS$_2$ channel using a sub-wavelength, nanopatterned hBN gate dielectric. Presence of strain in WS$_2$ has been verified through extensive spectroscopic characterization and that of strain-induced charge polarization through density functional theory calculations as well as piezo force microscopy. The pyrophototronic effect boosts dynamic photoresponsivity (0.7 A/W) and detectivity (1.2×10$^{15}$ Jones/cm) by up to 8x, and enhances the photodetection speed by 3x over the non-patterned (unstrained) phototransistor, demonstrating a path to ameliorating the responsivity-speed trade-off in 2D photodetectors. Analysis of gate voltage, wavelength and optical power dependence of the pyrophototronic current through measurements and band physics highlights its prominence under low channel population of electrostatically- or optically-induced free carriers. Gate tunability of the pyrophototronic current has been leveraged to design an optical spike-triggered dynamic accident alert system with speed-specific control for self-driving applications under low light conditions.

## 1 Introduction

In the modern era of artificial intelligence (AI), photodetectors are opening up exciting new opportunities for autonomous vehicles, intelligent surveillance, and industrial automation. Enhancing photodetector response and speed can significantly boost the performance of these AI-enabled applications by improving low-light detection,[1] data precision,[2] dynamic range,[3] and adaptability,[4] in addition to enabling real-time processing[5] and hyperspectral imaging.[6] However, a strong trade-off exists between the responsivity and speed of phototransistors since a longer carrier lifetime enhances the responsivity but decreases the speed of response.[7]

The pyrophototronic effect in non-centrosymmetric semiconductors offers a promising approach to address the aforementioned trade-off and simultaneously enhance both the photodetector responsivity and speed. Non-centrosymmetric semiconductors can generate a substantial dipole moment (μ) accompanied by non-zero ion charge polarization under applied strain, even without an applied electric field.[8–10] The polarization charge attracts free electrons to form polarization-bound stable states, and no current flows through the circuit in steady-state. If the surrounding temperature rises due to optical illumination, the dipoles lose their orientation due to increased thermal vibrations, decreasing μ and the polarization-bound charge. Bound charge carriers are released, resulting in current flow in the direction of an applied external electric field, thereby generating pyroelectricity through the pyrophototronic (PPE) effect.[11]

Metal oxides (ZnO,[12] $Ga_2O_3$,[13] SnO,[14] $Bi_2TeO_6$,[11] $VO_2$,[15]), compound semiconductors (CdS,[16] GaN,[17] SnS,[18] $Ag_2Se$,[19] $Ta_2NiS_5$[20]), ferroelectric[21] materials ($BiFeO_3$,[22] $BaTiO_3$,[23] P(VDF-TrFE[24])), and perovskites[25] exhibit such non-centrosymmetry-induced PPE. On the other hand, 2D layered materials hold significant promise for ultra-thin photonic and optoelectronic devices[26] due to the remarkable properties they exhibit:[27] a dangling-bond-free lattice, robust covalent in-plane bonding coupled with weak van der Waals out-of-plane interaction, exceptionally high carrier mobilities, layer-dependent electronic properties, and extensive surface areas for light absorption. Further, odd numbers of 2H-phase 2D transition metal dichalcogenide (TMDs) layers are known to be non-centrosymmetric due to the lack of an inversion center.[28–30] This has led to the demonstration of valley-selective circular dichroism,[31] large second-order nonlinear susceptibility,[32,33] corresponding to the optical second-harmonic generation (SHG) process[34] along with strain-induced piezoelectricity in monolayer as well as thicker, odd-numbered, TMD films.[35] These studies suggest that the PPE in strained, non-centrosymmetric, ion-polarized TMD films can be harnessed to improve 2D phototransistor performance metrics and their trade-off for future optoelectronic applications.

In addition to the lack of inversion symmetry in odd-numbered 2D TMD layers, they can endure substantial compressive as well as tensile strain due to their atomically thin nature[36] facilitating tuning of mechanical, optical, thermal, and electrical properties over a wide range. Several strain methods have been employed, including pre-strained elastomers/polymer substrates,[37] lattice or thermal expansion mismatch between the substrate and the 2D layers,[38] nanobubble or wrinkle formation,[39] scanning-probe-microscope-tip-induced deformation,[40] placement on patterned substrates,[41] etc. Most methods mentioned above, barring the use of patterned substrates, are not suitable for scaled-up technological applications due to the requirement of a continuous external strain source, or non-uniform strain distribution (elastomers/polymers substrates, scanning probe microscope tip, nanobubbles/wrinkles) or the release of local strain due to slippage between the substrate and the 2D film (lattice or thermal expansion mismatch). Unlike these techniques, patterned 2D dielectric substrates can induce uniform, stable, and conformal biaxial localized strain in 2D TMD films due to a nearly trap-free 2D-dielectric/2D-semiconductor[42] van der Waals interface.

In this work, we demonstrate the pyrophototronic effect in a nanopatterned 2D phototransistor ($T_P$) and its absence in a non-patterned control transistor ($T_{NP}$). For this purpose, a single non-centrosymmetric and ultra-thin (5-layer, 5L) WS$_2$ (2D TMD semiconductor) flake was exfoliated and transferred onto an hBN (2D dielectric) layer that was partly patterned to form T$_P$, and partly kept unpatterned to form T$_{NP}$. We show through extensive material characterization that nanopatterning of the hBN layer generates biaxial strain in WS$_2$, and through first principles density functional theory (DFT) calculations that the strain generates polarization charge in the non-centrosymmetric WS$_2$ flake. The enhanced light-matter interaction due to hBN nanopatterning improves photodetection, under steady-state illumination, by nearly 45x (from 0.35 A/W in T$_{NP}$ to 15.81 A/W A/W in T$_P$) and the detectivity by 125x. Importantly, though, dynamic switching of the incident light induces a time-dependent change in temperature of the polarized WS$_2$ layer, resulting in a four-stage pyrophototronic effect. PPE in patterned hBN/WS$_2$ enhances its dynamic photoresponsivity and detectivity by 8x, concurrently with a 3x increase in the speed of photoresponse (70 ms in T$_{NP}$, 20 ms in T$_P$) compared to the unstrained, non-patterned hBN/WS$_2$ phototransistor. Besides achieving an improved responsivity-speed trade-off, the electrostatic gate control of the pyrophototronic current in the ultra-thin WS$_2$ channel has been used to design a speed-dependent accident alert system that triggers a dynamic, event-based (sudden light change) alert to help avoid accidents in autonomous driving applications.

Below, under results and discussion, we first describe the phototransistor device structure and physical characterization results. Spectroscopic and microscopic characterization data of patterned and non-patterned WS$_2$, as shown in Figure 1, establish the presence of biaxial strain and strain-induced polarization charge in T$_P$. Figure 2 in the next sub-section confirms the presence of 5L WS$_2$ on patterned hBN using high-resolution transmission electron microscopy (HRTEM) and the effect of strain on in-plane polarization (electronic dipole moment) using first principles DFT calculations. It also shows improved steady-state photodetection in T$_P$ due to nanopatterning-led enhanced light-matter interaction. The subsequent sub-section on time-resolved photoresponse characterization, accompanied by Figure 3 and Figure 4(a), showcases primary experimental evidence of the pyrophototronic effect in T$_P$. Then, through electron energy band diagrams, as shown in Figure 4(b-e), we explain the origin of the pyrophototronic effect in T$_P$. The next sub-section presents detailed data and analysis for the incident wavelength-, power-, and the applied gate voltage-dependence of the pyrophototronic effect (Figure 5). Finally, Figure 6 depicts the design and operation of an accident alert system with speed-dependent control based on the 2D pyrophototronic transistor.

## 2 Results and Discussion

### 2.1 Phototransistor Device Structure and Physical Characterization

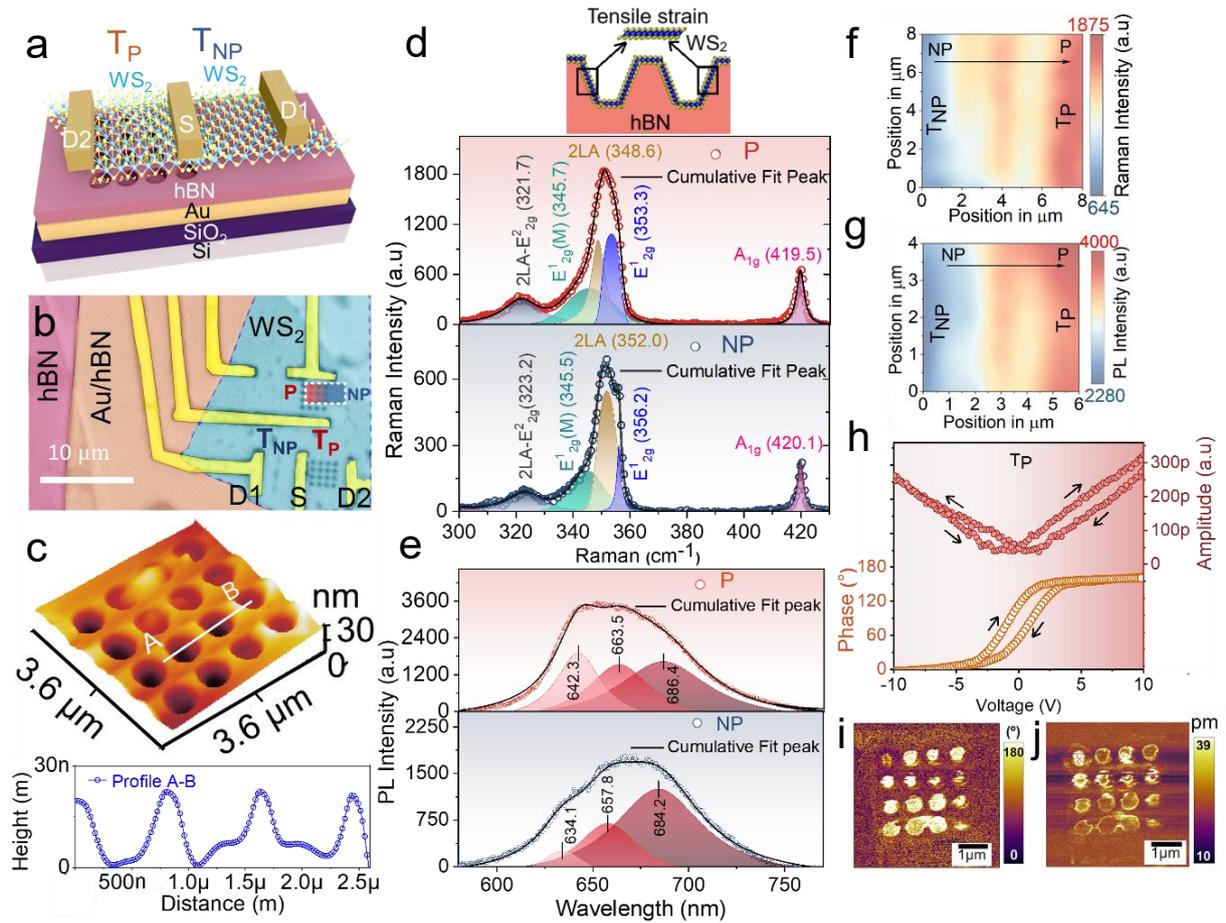

Figure 1. *Spectroscopic and microscopic characterization of biaxial strain and polarization charge in patterned ($T_P$) and non-patterned ($T_{NP}$) hBN dielectric-based 2D $WS_2$ Schottky phototransistors. (a) Device schematic, (b) optical microscope image, (c) AFM image of 5-layer $WS_2$ on patterned and non-patterned hBN dielectric. (d) Raman and (e) PL spectra under 532 nm illumination, (f) Raman mapping of $2LA+E^1_{2g}$ signal from patterned (P) and non-patterned (NP) regions of $hBN/WS_2$ devices under 532 nm laser illumination, (g) PL mapping of the 633 nm emission from patterned and non-patterned regions of $hBN/WS_2$ devices under 532 nm laser illumination, (h) phase and amplitude hysteresis loops measured on patterned $WS_2/hBN$ showing well-defined phase hysteresis (with a phase change of ~180°) and butterfly-shaped amplitude loops revealing the piezoelectric behavior of $hBN/WS_2$ due to hBN nanopatterning. (i) Piezoelectric domains in patterned hBN after application of −10 V and +10 V PFM tip bias, and (j) piezo-response amplitude images at $V_{ac}$=1 V with +10 V and −10 V dc voltages.*

Figures 1(a) and 1(b) show the schematic view and an optical microscope image of the two fabricated devices from the same $hBN/WS_2$ flake stack: $T_P$ and $T_{NP}$. These devices share common source (S) and gate (G) electrodes but different drain (D) contacts D1/D2. Atomic force microscope (AFM) image (Figure 1 (c)) of the patterned $hBN/WS_2$ region shows that the $WS_2$ is conformally placed on the etch-patterned hBN dielectric layer, thereby generating biaxial strain in $WS_2$.

As shown in Figure 1(d), Raman spectra collected from the patterned (P) and non-patterned (NP) regions (white dotted box in optical microscope image of Figure 1(b)) reveal interesting information regarding patterning-induced strain in the $WS_2$ film. The prominent Raman modes,[43] including the in-plane longitudinal optical phonon mode ($E^1_{2g}$) and the second-order acoustic mode 2LA at ~351 cm$^{-1}$, as well as the transverse optical phonon mode ($A_{1g}$) at ~419 cm$^{-1}$, are similar in both transistors $T_P$ and $T_{NP}$, except for the amplified Raman intensity observed in $T_P$. This can be attributed to the enhanced optical interference effect between the incident light and light scattered from the patterned hBN/WS$_2$ region, similar to the enhancement observed in a wrinkled WS$_2$ film as reported earlier.[37] However, the deconvoluted Raman spectra show that the $E^1_{2g}$ peak from NP (~356.2 cm$^{-1}$) is red-shifted by 3.5 cm$^{-1}$ from P (~353.3 cm$^{-1}$), indicating the presence of in-plane tensile strain in WS$_2$ from the patterned hBN dielectric as shown in the schematic in Figure 1(d). The $A_{1g}$ mode, which corresponds to the oscillation of sulfur atoms in the out-of-plane direction, is also red-shifted by 0.6 cm$^{-1}$ due to strain in the patterned hBN/WS$_2$ region. The larger red shift in $E^1_{2g}$ in-plane mode compared to the $A_{1g}$ out-of-plane peak indicates the predominant in-plane nature of the strain.[44]

PL spectra were also recorded to investigate the impact of biaxial strain on WS$_2$. The 5-layer WS$_2$ flake exhibits two prominent emission peaks associated with the direct and indirect band gap-related transitions at wavelengths of ~630 nm and ~830 nm, respectively[45] (Figure S2). The PL spectrum from P region shows enhancement in emission intensity and a blue shift in the peak position at 633 nm compared to the NP region. Deconvolution of the direct transition peak at 633 nm (Figure 1(e)), fitted with multiple Lorentz line shapes,[46] delineates the emission contributions from the A exciton, localized defect states at 650 nm, and defect-bound excitons at ~680 nm.[47] The PL peak corresponding to the direct band gap at 634.1 nm (1.95 eV) from NP shifts to 642.3 nm (1.93 eV) in P, indicating a 20 meV decrease in the direct band gap in the P region of WS$_2$, resulting from a strain-modified electronic band structure.[46,48] Biaxial tensile strain of 0.19% is estimated[49] for region P of the five layers WS$_2$, which results in broadening of the PL spectrum from the P region. Additionally, the peak position of the localized surface states shifts from 657.8 nm (1.88 eV) for NP to 663.5 nm (1.86 eV) for P. However, the energy difference between the localized surface states and the conduction band edge is 70 meV for P (1.95 eV – 1.88 eV= 70 meV) as well as NP (1.93 eV – 1.86 eV= 70 meV) region indicating that the tensile strain does not modify the energy distribution of the surface states with respect to the conduction band edge. The position and intensity of the PL peak corresponding to defect-bound excitons at 680 nm are similar for the P and NP regions. Based on finite difference time domain (FDTD) simulations, a high electric field enhancement factor of 6 is obtained for the electric field at the edge contour of a WS$_2$ flake transferred onto patterned hBN. The resultant enhanced coupling of incident light with sub-wavelength hBN patterning increases direct bandgap emission from strain-localized excitons at room temperature (Figure S3).

To further investigate the impact of biaxial strain on WS$_2$, Raman mapping was performed from the NP and P regions (Figure 1(f)) of the WS$_2$ flakes through precise x-y translational movement of the sample stage by 0.3 μm x-y increments while keeping the laser spot fixed. The Raman peak intensity is amplified in P for the in-plane longitudinal optical phonon mode ($E^1_{2g}$) and the second-order acoustic mode 2LA(M) at ~351 cm$^{-1}$. This can be attributed to the enhanced optical interference effect described earlier.[37] A PL spectral map (Figure 1(g)) also demonstrates an enhancement of PL intensity from the P region compared to NP at the direct band gap wavelength of ~633 nm.

As shown in Figure 1(h), the biaxial strain-induced piezoelectricity of WS$_2$ was verified by piezo-response force microscopy (DART-PFM, Asylum Research) under resonance-enhanced mode[50] at room temperature under natural ambient conditions. A triangular saw-tooth voltage waveform (0.1 – 0.2 Hz) was applied at the conductive tip to switch the polarization and obtain a hysteresis loop. The amplitude-bias curve in Figure 1(h) shows a V-shaped butterfly loop, and the phase curve shows hysteresis behaviour, confirming the presence of biaxial strain-induced polarization in the patterned WS$_2$/hBN stack, unlike the non-patterned region. The piezo phase map shown in Figure 1(i) indicates that the patterned region exhibits enhanced piezo-response compared to the non-patterned region due to the presence of in-plane polarization charge.[45] The phase and amplitude maps show that the response follows the AFM topography of Figure 1(c), and the piezo-response appears at the edge contours of the hBN/WS$_2$ patterns. However, in amplitude imaging, piezo-response gradually decays from the edge to the center of the patterns, indicating maximum strain at the pattern wall that gradually decays towards the flat hBN surface at the center. We have theoretically validated the experimental observation of charge polarization in WS$_2$ due to pattern-induced biaxial strain as detailed in the next sub-section 2.2.

**2.2 TEM Imaging and Density Functional Theory Calculations**

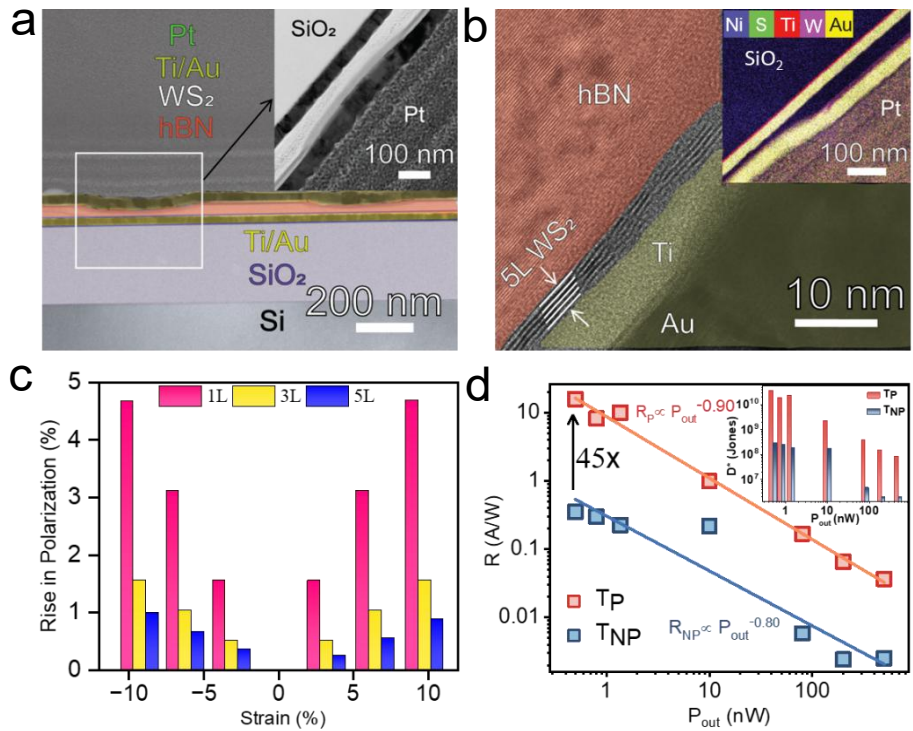

*Figure 2. (a) Low magnification cross-sectional HRTEM image of device T$_P$ showing two patterned hBN regions, inset shows the high magnification image of distinct hBN and WS$_2$ layers in one such patterned region of the*

*device, (b) 5-layer WS$_2$ on hBN and inset shows the HAADF-STEM elemental mapping of T$_P$ confirming the presence of constituent elements of WS$_2$, (c) DFT-based calculations of the % increase in 2D dipole moment as a function of compressive and tensile biaxial strain for 1 layer (1L), 3 layer (3L), and 5-layer (5L) WS$_2$, (d) power-dependent steady-state photoresponsivity of T$_P$ and T$_{NP}$ at V$_{gs}$= -3 V and V$_d$= 1 V for 660 nm illumination, inset shows the power-dependent steady-state specific detectivity of T$_P$ and T$_{NP}$ at V$_{gs}$= -3 V and V$_d$= 1V.*

Bulk WS$_2$, being centrosymmetric, has zero net polarization due to a balanced distribution of positive and negative charge. On the other hand, non-centrosymmetric odd number of WS$_2$ layers can acquire finite charge polarization under structural distortion introduced by volumetric or biaxial strain. We first performed HRTEM characterization to determine the layer count in WS$_2$ of T$_P$ and T$_{NP}$ and to investigate the presence of interfacial strain between the 2D layers. Figure 2(a) depicts a cross-sectional HRTEM image of device T$_P$ with two patterned hBN regions. The inset magnifies one such hBN patterned region, showing a WS$_2$/Ti/Au/Pt stack transferred and conformally deposited on top of the hBN patterned region. Further, Figure 2(b) shows a clean and high-quality van der Waals interface between the multilayered patterned hBN layer and a continuous 5L WS$_2$ film. It also depicts the partial lattice deformation of WS$_2$ in the patterned hBN region arising from tensile strain, as also inferred from the Raman spectra (Figure 1(d)). The continuity of the WS$_2$ lattice in Figure 2(b) indicates the absence of any strain-induced cracks in the WS$_2$ film and at the WS$_2$/hBN interface[51]. Inset of Figure 2(b) shows elemental mapping and intensity using high-angle annular dark field scanning transmission electron microscopy (HAADF-STEM) and energy-dispersive x-ray spectroscopy (EDS), indicating discrete and clean Au (gate)/hBN (dielectric) and hBN/WS$_2$ interfaces. Having established the non-centrosymmetric nature of the WS$_2$ flake, we performed DFT calculations of the % increase in 2D dipole moment as a function of 2D biaxial strain for 1-layer (1L), 3-layer (3L), and 5-layer (5L) WS$_2$. To avoid any interference from periodic replication, the slabs of 1L, 3L, and 5L WS$_2$ were quantum confined within a vacuum of ~12 Å along the c-axis. The three cases were investigated under biaxial tensile (positive) and compressive (negative) strain of up to 10%, and the % increase in resultant 2D dipole moments is summarized in Fig. 2(c). Both kinds of strain result in a substantial increase in charge polarization that increases with decreasing number of layers, as expected. Details of the theoretical calculations are given in Figures S4-S7 and the associated text. In addition, for 1L WS$_2$, we have also investigated the impact of both compressive and expansive volumetric strain, and the results are presented in Figure S7. Expansive volumetric strain increases polarization, whereas volumetric compression decreases it, and the system undergoes a direct-to-indirect crossover. In summary, the theoretical calculations show that structural deformations caused by biaxial or volumetric strain induce a charge imbalance within the WS$_2$ system, resulting in non-zero piezoelectric polarization.

Next, we examined the electrical performance of T$_P$ and T$_{NP}$ transistors in dark (Figure S8 for dark output characteristics) and under steady-state 660 nm laser illumination (Figure S9 and associated text for transfer characteristics under dark and illumination). Figure 2(d) depicts the photoresponsivity, $R(\lambda) = \frac{I_{ph}}{P_{opt}}$ where P$_{opt}$ is the optical power at wavelength λ and I$_{ph}$ is the photocurrent (I$_{ph}$= I$_{light}$-I$_{dark}$ where $I_{dark}$ and $I_{light}$ are the drain currents under dark and light exposure respectively) of $T_P$ and $T_{NP}$ for varying light illumination power at fixed $V_{gs}$= -3 V, $V_d$= 1 V and $\lambda$= 660 nm. $T_P$ and $T_{NP}$ show a peak responsivity of 15.8 A/W and 0.35 A/W at 500 pW power, respectively. Peak R($\lambda$) of $T_P$ is thus ~45X higher than that of $T_{NP}$, which can be attributed to the slightly lower $I_{dark}$ but predominantly higher $I_{light}$ as shown in supporting Figure S9(a). The enhancement of $I_{light}$ and R($\lambda$) in $T_P$ and $T_{NP}$ can also be corroborated with enhanced PL intensity from $T_P$ compared to $T_{NP}$ (Figure 1(e)) and a PL enhancement factor (EF) of 6 for $T_P$ with respect to $T_{NP}$, as obtained from FDTD simulations in Figure S3. Shot noise-limited specific detectivity (D*($\lambda$)= $\frac{R(\lambda) \times \sqrt{A}}{\sqrt{2qI_{dark}}}$)[7], where $q$ is the electron charge and A is the active area of the device, is plotted vs incident light power at $\lambda$= 660 nm in the inset of Figure 2(d). Peak specific detectivity is estimated to be 3.54×10^{10} Jones and 2.8×10^8 Jones at fixed $V_{gs}$= -3 V, $V_d$= 1 V, and $\lambda$= 600 nm for $T_P$ and $T_{NP}$, respectively. The higher R and lower $I_{dark}$ for $T_P$ results in a 125x larger D* compared to $T_{NP}$ near the off-state ($V_{gs}$= -3 V). In summary, $T_P$ shows improved steady-state photodetection performance as compared to $T_{NP}$, primarily due to enhanced light-matter interaction.

## 2.3 Time-resolved Photoresponse: Evidence of Pyrophototronic Effect

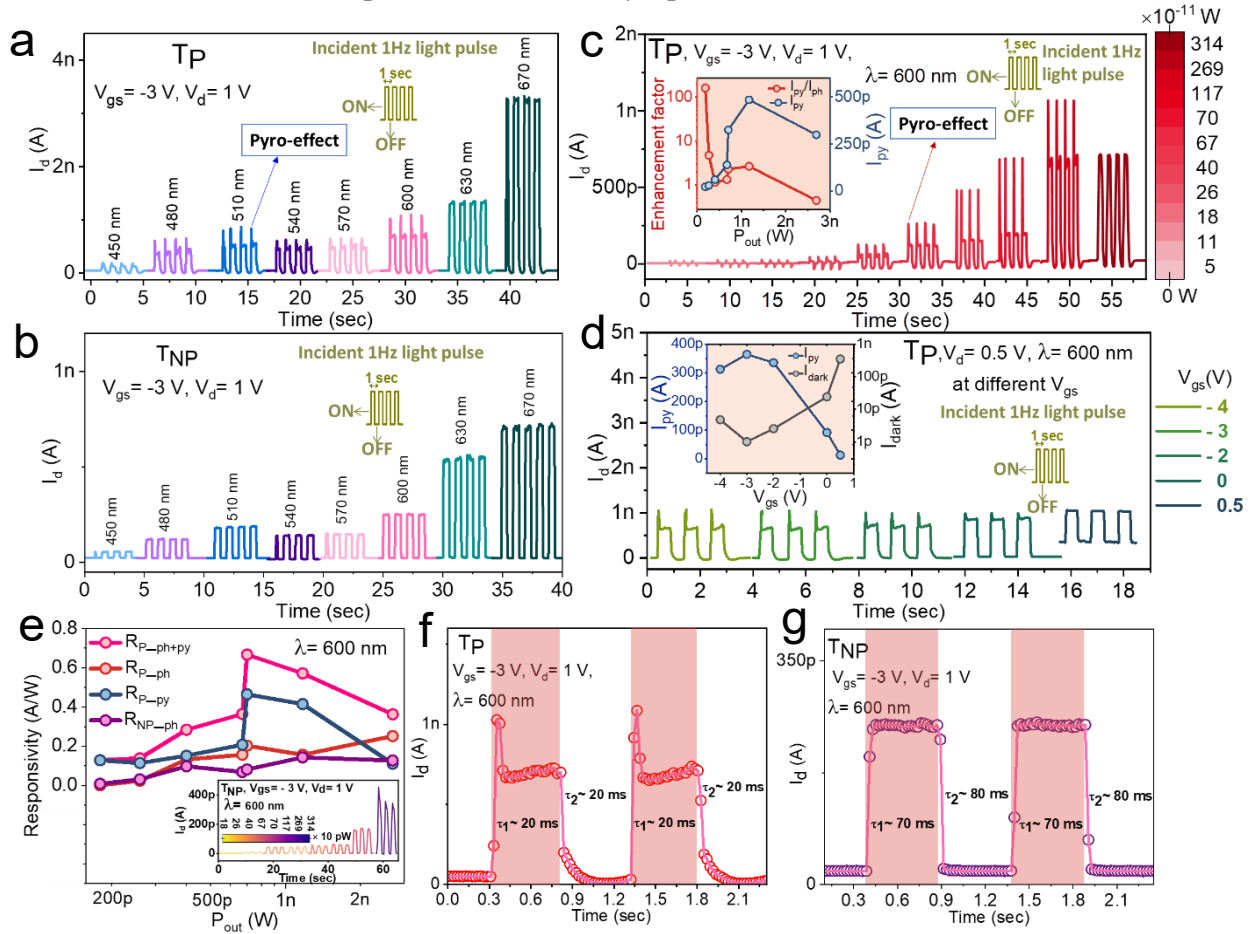

Figure 3. *Time-resolved (1 Hz optical pulse) dynamic optoelectronic characterization of $T_P$ and $T_{NP}$ phototransistors. Time-resolved photoresponse ($I_d$ vs time) plots for devices (a) $T_P$ and (b) $T_{NP}$ at different wavelengths of light under -3 V gate-to-source bias and 1 V drain-to-source bias, and 1 nW*

*laser power. The patterned device shows pyrophototronic-enhanced photoresponse. (c) Optical power-dependent time-resolved pyrophototronic photoresponse ($I_d$ vs time) plot for $T_P$. The inset shows the enhancement factor ($I_{py}$:$I_{ph}$) and $I_{py}$ vs laser power. (d) $V_{gs}$-dependent time-resolved pyrophototronic photoresponse ($I_d$ vs time) plot for $T_P$. The inset shows the variation of extracted $I_{py}$ and $I_{dark}$ with $V_{gs}$. (e) Comparison of optical power-dependent pyro+photoresponsivity ($R_{p\_ph+py}$), photoresponsivity ($R_{p\_ph}$), pyroresponsivity ($R_{p\_py}$) of $T_P$ with the photoresponsivity ($R_{NP\_ph}$) of $T_{NP}$. Rise and fall times of (f) $T_P$, and (g) $T_{NP}$ under 1 Hz, 1 nW optical pulse illumination.*

This sub-section reports direct experimental evidence of the pyrophototronic effect in the patterned transistor $T_P$. The transient current $I_d$ measured for illumination at different wavelengths (450 nm to 670 nm) is shown for $T_P$ (Figure 3(a)) and $T_{NP}$ (Figure 3(b)). Figure 3(a) shows the presence of a distinguishable and sharp transient peak corresponding to the pyrocurrent $I_{py}$, in addition to the photocurrent $I_{ph}$ (above the base $I_{dark}$), for device $T_P$ in I-t curves at the instants where the incident light (from 450 nm to 600 nm wavelength) was turned on or off. This is the signature of the pyrophototronic effect.[52,53] It can be attributed to the biaxial tensile strain-induced polarization charge in $WS_2$ and explained in detail through energy band diagrams later in this manuscript. In contrast, as shown in Figure 3(b), the dynamic photoresponse of device $T_{NP}$ indicates the presence of only conventional $I_{ph}$ under light on/off instants; the absence of a transient pyrocurrent peak is due to the lack of significant strain-induced polarization charge. Figure 3(c) shows the illumination power-dependent generation of $I_{py}+I_{ph}$ and $I_{ph}$ at $V_{gs}$= -3 V and $V_d$= 1 V for a fixed wavelength of 600 nm. $I_{py}$ increases monotonically with the incident laser power ($P_{opt}$ varying from 50 pW to 3.1 nW) upto 1.2 nW. $I_{ph}$, on the other hand, is nearly zero for extremely low powers (below 180 pW) and then increases monotonically with incident $P_{opt}$. The transient current ($I_{py}+I_{ph}$) exhibits a maximum enhancement factor $I_{py}$:$I_{ph}$ =100 at a low laser power of approximately 300 pW, as shown in the inset of Figure 3(c). Figure 3(d) shows the $V_{gs}$-dependent ($V_{gs}$ varying from n-type to p-type state of $T_P$ → 0.5 V to -4.5 V) transient photoresponse at fixed $V_d$= 1 V, λ=600 nm, $P_{opt}$= 1 nW. The device exhibits an enhanced pyrophototronic current in the low dark current regime ($V_{gs}$= -3 V in the near-off condition of the p-type branch). We explain this particular phenomenon further in section 2.4. At more negative gate voltage, i.e., $V_{gs}$ much less than $V_{th}$ for the p-type branch, the dark current increases, and the transient $I_{py}$ current reduces due to the trade-off between the pyrophototronic effect and bias-dependent Joule heating.[54] Likewise, $I_{py}$ nearly vanishes for $V_{gs}$= 0.5 V when the n-type dark current is large. The inset depicts the trends in extracted $I_{py}$ and $I_{dark}$ vs $V_{gs}$, $I_{py}$ decreases for higher $I_{dark}$ and vice versa.

Figure 3 (e) shows photoresponsivity (R(λ)) vs incident light power for $T_{NP}$ and $T_P$ at $V_{gs}$= -3 V and $V_d$ = 1 V for the transient photoresponse at λ= 600 nm. The inset shows the transient photoresponse of $T_{NP}$ for varying power, similar to the data for $T_P$ shown in Figure 3(c). It is important to note that the range of powers (180 pW to 2 nW) has been chosen to highlight the effect of $I_{py}$, which is dominant at low power illumination. Further, R(λ) for $T_P$ has been broken down into constituent components resulting from $I_{py}$ and $I_{ph}$, showing that the enhancement in R(λ) for $T_P$ is primarily driven by the appearance of $I_{py}$. A peak enhancement of 8x is seen at 700 pW of incident power. Similarly, detectivity $(D(\lambda)= \frac{R(\lambda)}{\sqrt{2qI_{dark}}})$ and specific detectivity $(D^*(\lambda)= \frac{R(\lambda)\sqrt{A}}{\sqrt{2qI_{dark}}})$, along with their components resulting from $I_{py}$ and $I_{ph}$, are shown in Figure S10. The rise and decay times of dynamic photoresponse have also significantly improved in device $T_P$ (Figure 3(f)) compared to device $T_{NP}$ (Figure 3(g)) due to the polarization charge-driven pyrophototronic effect (as explained next through Figure 4(b-c)). This shows the benefit of the pyrophototronic effect in improving the speed-responsivity trade-off in 2D photodetectors.

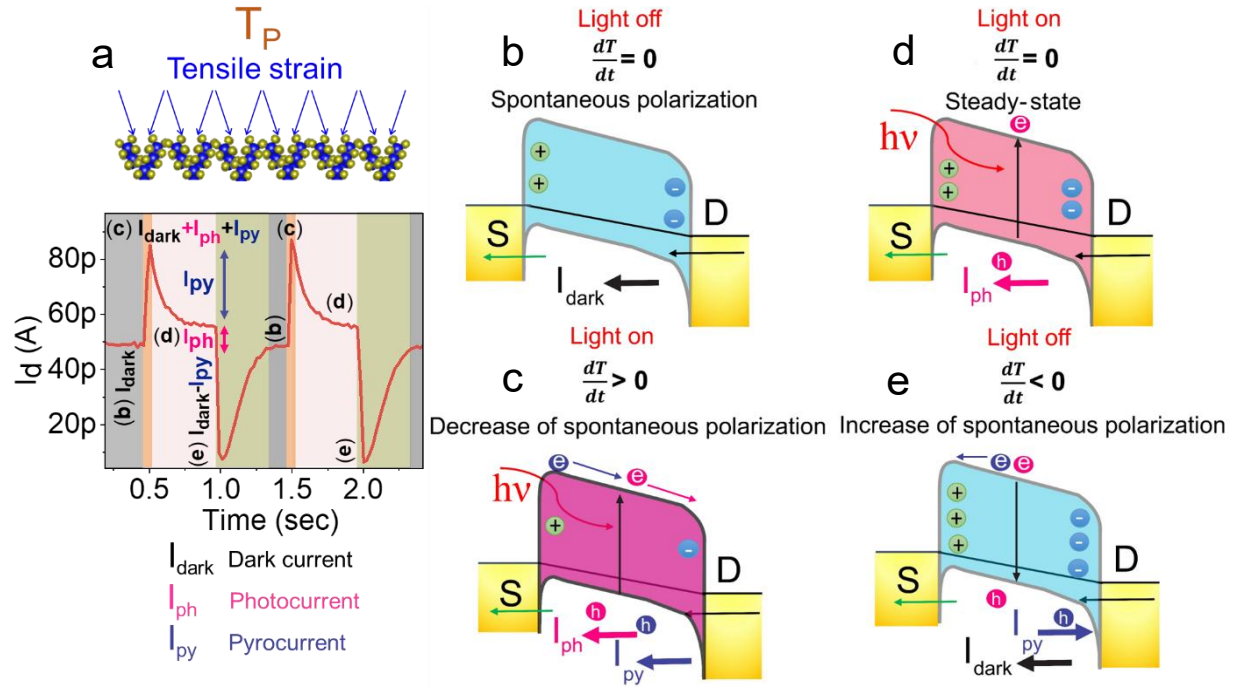

Figure 4. *Mechanism of the pyrophototronic effect in the strained WS$_2$ phototransistor $T_P$. (a) Detailed time-resolved response of device $T_P$ under 600 nm wavelength laser light illumination (power: 180 pW, frequency of turning light on and off = 1 Hz) at $V_{gs}$= -3 V and $V_d$= 1 V, highlighting the various current components and segments that are explained through energy band diagrams. Energy band diagrams which show: (b) strain-induced spontaneous polarization in the absence of light, (c) local heating and increased thermal vibration under light exposure reduce spontaneous polarization to release free carriers ($I_{py}$), (d) spontaneous polarization is regained under prolonged light exposure as temperature stabilizes, (e) local cooling when light is turned off leads to an increase in spontaneous polarization and free carrier current ($I_{py}$) that opposes $I_{dark}$.*

We now explain the mechanism of the pyrophototronic effect in detail with the help of energy band diagrams. Figure 4(a) shows the detailed time-resolved photoresponse of $T_P$, highlighting the various components and segments of the current response, under 600 nm light illumination at $V_{gs}$= -3 V and $V_d$= 1 V. The different stages of the pyrophototronic effect have been marked as (b-e), and corresponding band diagrams are shown in Figure 4(b-e). Figure 4(b) shows the under-dark condition where dark hole current $I_{dark}$ flows from the drain to the source terminal of the WS$_2$ phototransistor. Nanopatterning-induced biaxial tensile strain in WS$_2$ of device $T_P$ results in charge polarization (Figure 1(h)) and the formation of a dipole moment, $\mu$= q × d, where q is the dipole electric charge, d is the distance between the charge centroids, and the dipole moment per unit volume is the net polarization. The biaxial strain in $WS_2$ generates a significant µ and a non-zero polarization without an electric field. The polarization charge attracts nearby free carriers to form polarization-bound stable states that do not affect $I_{dark}$. Under illumination (Figure 4(c)), $WS_2$ absorbs light and generates electron-hole pairs, which are driven by the drain-induced electric field, producing photocurrent $I_{ph}$. Additionally, the incident photons generate a temperature change (dT/dt > 0); the charge dipoles lose orientation due to increased thermal vibration, and spontaneous polarization decreases. The polarization-bound charge also decreases, releasing free carriers that redistribute and flow under the influence of the electric field, thereby generating $I_{py}$. As the temperature stabilizes due to continuous light exposure (Figure 4(d)), dT/dt is reduced to zero, $I_{py}$ becomes zero, leaving behind only the photocurrent $I_{ph}$ and stable spontaneous polarization. When light is turned off (Figure 4(e)), temperature reduces, the material cools (dT/dt < 0), and spontaneous polarization increases. The current $I_{py}$ now flows in the reverse direction as the charge carriers move to reform and increase the polarization-bound states. Under prolonged light-off conditions, the temperature again stabilizes (dT/dt= 0), charge-neutral conditions with stable spontaneous polarization are restored, and the current returns back to $I_{dark}$ as shown in Figure 4(b).

## 2.4 Dependence of Pyrophototronic Effect on Gate Voltage and Incident Light Wavelength and Power

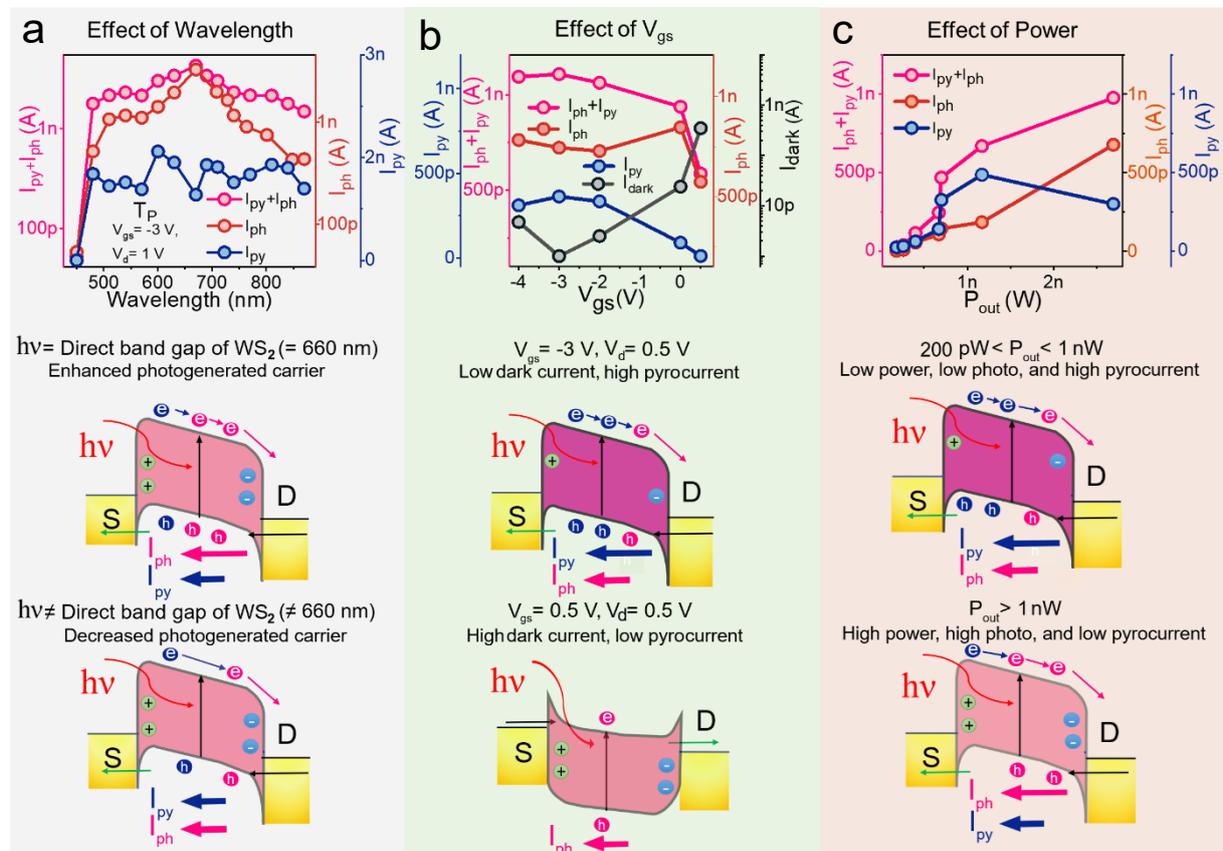

Figure 5. *Pyro+photo current ($I_{py}+I_{ph}$), photocurrent ($I_{ph}$), and pyrocurrent $I_{py}$ of $T_P$ (a) under different wavelengths of light illumination at $V_{gs}$= -3 V and $V_d$= 1 V for laser power of 1 nW, (b) under different*

*$V_{gs}$ at $V_d$= 0.5 V at 600 nm wavelength, 1 nW laser illumination, and (c) under different laser powers for 600 nm wavelength illumination at $V_{gs}$= -3 V and $V_d$= 1 V.*

Finally, we explain the dependence of the pyrophototronic effect on the incident wavelength and power, as well as the gate voltage applied to T$_P$. Figure 5(a) shows $I_{py}$+$I_{ph}$, and the constituent $I_{ph}$ and $I_{py}$, vs wavelength (450–900 nm) at $V_{gs}$= -3 V and $P_{opt}$= 1 nW. T$_P$ exhibits a visible-to-near-infrared photoresponse. In the 450 – 600 nm range, the $I_{py}$+$I_{ph}$ and $I_{ph}$ increase gradually since the power is fixed, and hence the number of incident photons increases with wavelength. At 660 nm, these responses converge, after which $I_{ph}$ decreases due to sub-bandgap illumination. Wavelength of 660 nm corresponds to near-band-edge absorption of WS$_2$, resulting in high absorption and photoresponse ($I_{ph}/I_{dark}$~$10^5$), while the $I_{py}$+$I_{ph}$ response is not prominent.[55] It is interesting to note that $I_{py}$ is nearly independent of wavelength. The band diagrams illustrate the carrier action and $I_{ph}$ and $I_{py}$ components at illumination of 660 nm and otherwise.

The effect of different $V_{gs}$ values on the pyrophototronic effect is explained next for a fixed wavelength of 600 nm at 1 nW power. Since the gate voltage affects the dark current, $I_{dark}$ is also included in the plot along with $I_{ph}$+$I_{py}$, $I_{ph}$, and $I_{py}$ for different $V_{gs}$ values, ranging from -4.5 V to 0.5 V, as shown in Figure 5(b). For $V_{gs}$= -4.5 V to 0 V, the pyrophototronic effect is observed, $I_{py}$ is detectable, and $I_{dark}$ is <100 pA. However, for $V_{gs}$= 0.5 V, $I_{dark}$ increases substantially, and $I_{py}$ disappears due to compensation of pyrophototronic effect through Joule heating.[55] In general, throughout the $V_{gs}$ range, $I_{py}$ increases with decreasing $I_{dark}$ and vice versa. $I_{ph}$, on the other hand, shows a weaker dependence on $I_{dark}$. The band diagrams illustrate carrier action at $V_{gs}$= -3 V (off, low $I_{dark}$) and $V_{gs}$= 0.5 V (on, high $I_{dark}$).

Figure 5(c) shows $I_{ph}$+$I_{py}$, $I_{ph}$, and $I_{py}$ for varying illumination powers ranging from 50 pW to 2.7 nW, while the wavelength remains fixed at 600 nm. $I_{py}$ peaks at 1.2 nW, decreases thereafter, and disappears above 3 nW (similar to Figure 3(c)), while $I_{ph}$ and consequently $I_{ph}$+$I_{py}$ increase monotonically with incident power. At low illumination power, 180 pW to 1 nW, the device exhibits significant $I_{py}$. In this regime, photo-generated charge carriers are limited, and pyro-related charge carriers dominate the four-stage pyrophototronic response. This suggests that biaxial strain-related polarization charge in WS$_2$ is highly temperature-sensitive and can respond at very low light intensities. However, at high illumination power, WS$_2$ absorbs a significant amount of light, leading to a large number of photo-generated carriers, which dominate the overall current transport. In summary, Figure 5 shows that when the channel is less populated with electrostatically- or optically-induced free carriers, the pyrophototronic effect is more prominent.[56] This study demonstrates the interplay between $I_{dark}$, $I_{py}$, and $I_{ph}$ in the device and could be beneficial for applications requiring weak light detection beyond the band edge absorption of the photodetector channel material.

## 2.5 A Pyrophototronic Accident Alert System

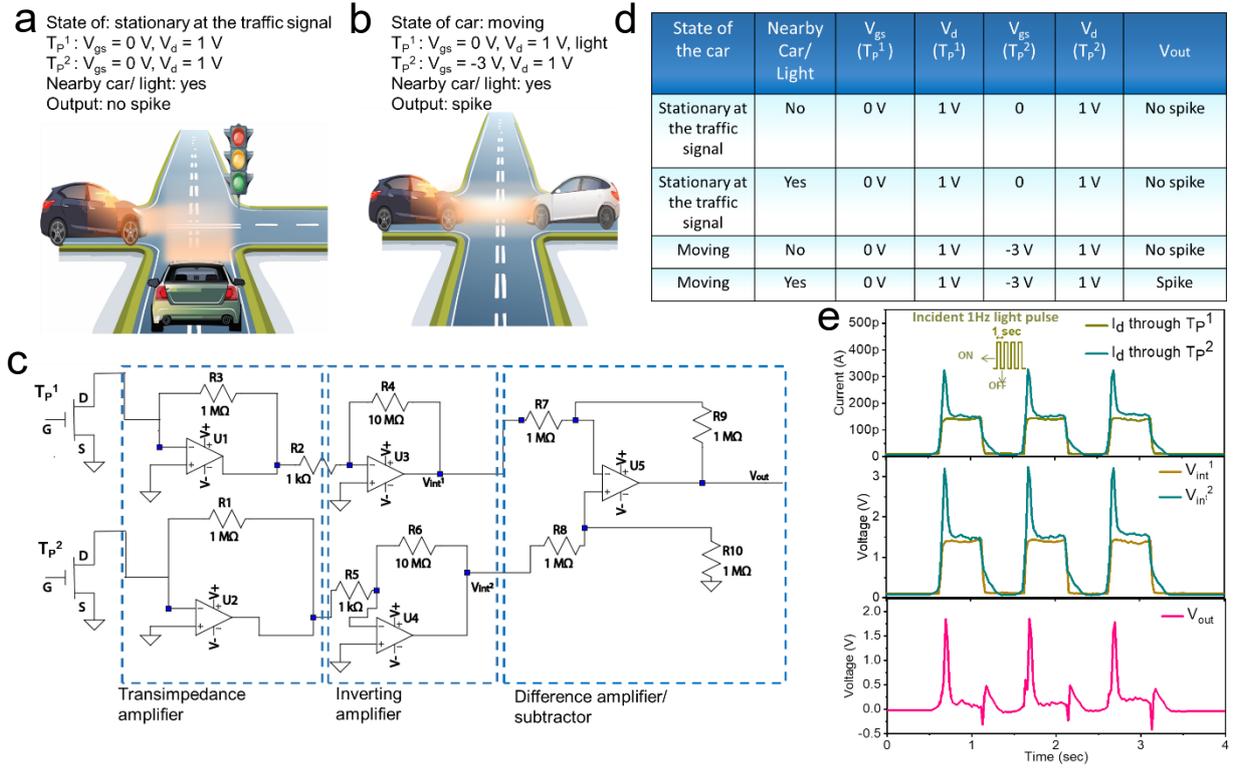

Figure 6. *Pyrophototronic accident alert system using a differential configuration of two patterned phototransistors $T_P^1$ and $T_P^2$: (a) Schematic of a stationary car ($V_{gs}$= 0 V for both transistors) at a traffic signal such that incoming light due to a nearby car/traffic signal only generates photocurrent ($I_{ph}$) and no pyrocurrent ($I_{py}$) in either $T_P^1$ or $T_P^2$. The differential circuit (shown in Figure 6(c)) subtracts the similar photocurrents through $T_P^1$ and $T_P^2$, does not generate a spike, and hence no alarm is activated in this case (sleep mode). (b) Schematic of a moving car ($V_{gs}$= 0 V for $T_P^1$, $V_{gs}$= -3 V for $T_P^2$). Incoming light from another car generates pyro+photocurrent in $T_P^2$ ($V_{gs}$= -3 V) and only photocurrent in $T_P^1$ ($V_{gs}$= 0 V). The circuit generates spikes in this case, triggering an alarm/alert (trigger mode) for the car driver to take necessary action and avoid an accident. (c) A differential circuit to subtract the currents through the two phototransistors (after converting the currents to voltage and amplification). (d) The input-output table captures the alarm system described above. (e) SPICE simulated current waveforms through $T_P^1$ and $T_P^2$ for a moving car with light on and off, corresponding intermediate circuit voltage waveforms ($V_{int}^1$ and $V_{int}^2$), and the difference between these two voltage waveforms ($V_{out}$), as obtained through the difference amplifier stage of (c).*

As a practical utilization of the pyrophototronic transistor, we demonstrate an accident alert/alarm system built using two patterned transistors $T_P^1$ and $T_P^2$, which are photo-sensitive ($V_{gs}$= 0 V for both) and can be selectively tuned through gate voltage to become pyro-sensitive ($V_{gs}$= 0 V for $T_P^1$, $V_{gs}$= -3 V for $T_P^2$) as described earlier in this manuscript. Figure 6(a) shows the schematic of a black, stationary car at a traffic signal. Low or zero speed sets the alert system in sleep mode ($V_{gs}$= 0 V for both transistors). Incoming light through another car, or the traffic signal, generates only photocurrent ($I_{ph}$) and no pyrocurrent ($I_{py}$) in either $T_P^1$ or $T_P^2$ because we have set the control voltage $V_{gs}$= 0 V for both $T_P^1$ and $T_P^2$. Figure 6(b) shows the black car moving, and its speed beyond a certain threshold sets the alarm system in trigger mode ($V_{gs}$= 0 V for $T_P^1$, $V_{gs}$= -3 V for $T_P^2$). In that case, incoming light from another car (white in Figure 6(b)) generates $I_{ph}+I_{py}$ current in $T_P^2$ but only $I_{ph}$ in $T_P^1$. The alert system has been designed using a LTspice-based differential circuit (shown in Figure 6(c)), which subtracts the drain current ($I_d$) through $T_P{}^1$ from that through $T_P{}^2$ using a difference amplifier/subtractor stage after first converting them into voltages through a trans-impedance amplifier stage and subsequently amplifying them to $V_{int}{}^1$ and $V_{int}{}^2$ through an inverting amplifier stage. $V_{out}$ is the final output voltage that spikes (proportional to the difference in $I_{ph}+I_{py}$ and $I_{ph}$) if the moving car suddenly sees incoming light from another vehicle, or does not spike (difference in $I_{ph}$ of the two transistors being nearly zero). These input and output conditions are highlighted in the table of Figure 6(d). The corresponding waveforms for $V_{int}{}^1$, $V_{int}{}^2$, and $V_{out}$, as obtained from SPICE simulations of the circuit in Figure 6(c), are shown in Figure 6(e). The nature of the pyrophototronic effect is particularly suited for accident alert because it captures sudden, dynamic changes in light intensity, works well in accident-prone dark environments, and is highly sensitive to low optical powers, thereby triggering alarms even for far-away or weak changes in light intensity.

## 3 Conclusions

**Table 1. Summary and comparison of key performance parameters for various pyrophototronic effect-enhanced photodetectors reported in recent literature.**

| Material(s) | Dimensionality and 2D thickness | Spectral range (nm) | Response time (ms) @λ (nm) | Peak responsivity (A/W) @λ (nm) | Peak specific detectivity $D^*(\lambda) = \frac{R(\lambda)\sqrt{A}}{\sqrt{2qI_{dark}}}$ (Jones) @λ (nm) | Peak detectivity $D(\lambda) = \frac{R(\lambda)}{\sqrt{2qI_{dark}}}$ (Jones/cm) @λ (nm) | Phototransistor/gate tunability | Application |
|---|---|---|---|---|---|---|---|---|
| **ITO/perovskite/PTAA: P3HT/Au[57]** | 2D (100 nm)/3D | 360–785 | 0.08 @320 | 2.8 @320 | $2.5 \times 10^{20}$ @320 | ------- | N/N | Optical Communication |
| **Cs$_2$PbI$_2$Cl$_2$/PCBM/BCP/Bi$_2$TeO$_6$/CsSn$_{0.5}$Pb$_{0.5}$IBr$_2$[11]** | 3D/2D (7.1 nm) | 450–950 | 0.02 @530 | 1.46 @530 | $3.44 \times 10^{13}$ @530 | ------- | N/N | Triple-Encryption Imaging Sensing System |
| **HfO$_2$/NiO/ZnO event-based sensor[58]** | 3D/3D | 400–808 | 0.05 @380 | $6 \times 10^{-3}$ @380 | ------- | ------- | N/N | Simplified Broadband Event-based Sensor |
| **Al/Si/SiOx/ZrO$_2$/ITO devices[59]** | 3D/3D | 405–1064 | 0.002 @940 | 3.4 @940 | $1.2 \times 10^{10}$ @940 | ------- | N/N | CMOS-Compatible Near-Infrared Photodetectors |
| **Ga$_2$O$_3$/SiO$_2$/Si[60]** | 3D/3D | 940 | 0.0004 @940 | $1 \times 10^{-5}$ @940 | $1.24 \times 10^{9}$ @940 | ------- | N/N | In-sensor Spatiotemporal Parallel Optical Information Processing |
| **GaON/GaN[61]** | 3D/3D | 255–365 | 0.008 @365 | $17.1 \times 10^{-3}$ @365 | $2.5 \times 10^{8}$ @365 | ------- | N/N | Electrolyte-assisted Inverse Encrypted Optical Communication System |
| **Ag/(HDA)BiI$_5$/Ag[62]** | 3D/3D | X-ray, 377–980 | ------- | ------- | $1.84 \times 10^{8}$ @520 | ------- | N/N | N |
| **Ag/(BA)$_2$SbBr$_5$/Ag[63]** | 3D/3D | 377–980 | 0.09 | ------- | ------- | ------- | N/N | N |
| **SnS/Si[64]** | 3D/3D | 365–850 | 0.012 @750 | 0.013 @750 | ------- | $3 \times 10^{14}$ @750 | N/N | N |

| Material(s) | Dimensionality and 2D thickness | Spectral range (nm) | Response time (ms) @λ (nm) | Peak responsivity (A/W) @λ (nm) | Peak specific detectivity $D^*(\lambda) = \frac{R(\lambda)\sqrt{A}}{\sqrt{2qI_{dark}}}$ (Jones) @λ (nm) | Peak detectivity $D(\lambda) = \frac{R(\lambda)}{\sqrt{2qI_{dark}}}$ (Jones/cm) @λ (nm) | Phototransistor/gate tunability | Application |
|---|---|---|---|---|---|---|---|---|
| Ag/ZnO Nanoparticles[65] | Nanoparticle/3D | 325 | 25.87 @325 | $8.82 \times 10^{-5}$ @325 | $4.9 \times 10^{10}$ @325 | ------- | N/N | N |
| Water/Si[66] | Liquid/3D | 400–800 | 20 @400 | 0.754 @400 | ------- | $1.33 \times 10^{13}$ @400 | N/N | N |
| MoS$_2$-water[67] | 2D (bulk MoS$_2$ crystal)/liquid | 470 | 3.2 @470 | $24.6 \times 10^{-3}$ @470 | $2.85 \times 10^{8}$ | ------- | N/N | N |
| p-Si/n-BaTiO$_3$[68] | 3D/3D | 365–632 | 37 | $1.34 \times 10^{-3}$ @365 | $2.25 \times 10^{10}$ @365 | ------- | N/N | N |
| WSe$_2$/Ta$_2$NiS$_5$ nanoflakes[20] | 2D/2D (30 nm/36 nm) | 405–1064 | 17 @660 | $121 \times 10^{-3}$ @660 | $2 \times 10^{12}$ @660 | ------- | N/N | N |
| Patterned hBN dielectric/WS$_2$ | 2D (45 nm)/2D (3 nm) | 450–870 | 20 @600 | 0.7 @600 | $4.7 \times 10^{10}$ @600 | $1.2 \times 10^{15}$ @600 | Y/Y | Speed-dependent Accident Alert System |

The table above summarizes and compares key features and parameters of recently reported pyrophototronic effect-enhanced photodetectors. This work has demonstrated it through strain-engineering of an ultra-thin, non-centrosymmetric 5-layer WS$_2$ phototransistor channel using a nanopatterned hBN dielectric. The presence of biaxial tensile strain has been confirmed through extensive PL, Raman, as well as HRTEM characterization, and the polarization charge has been validated through DFT calculations and PFM analysis. The pyro-photoresponse manifests clearly in time-resolved optical measurements and has been measured and analyzed for its wavelength, optical power, and gate voltage dependence. In this regard, this work reports the first demonstration of electrostatically tunable pyrophototronic effect, made possible due to the ultra-thin 2D semiconducting WS$_2$ film used as the charge-polarized, photoresponsive transistor channel material. The pyro-photoresponse spans a broad spectral range (450-870 nm) with significantly higher dynamic photoresponsivity (0.7 A/W at 600 nm) and detectivity ($1.2 \times 10^{15}$ Jones/cm, specific detectivity of $4.7 \times 10^{10}$ Jones) values than recent reports. It ameliorates the photoresponsivity-speed trade-off that limits the performance of conventional 2D photodetectors by concurrently enhancing dynamic photoresponsivity (switching speed) by 8x (3x). Finally, we exploit its gate tunability to demonstrate a speed-specific accident alert system that also leverages the dynamic nature of the pyrophototronic effect and its effectiveness in dark and low light intensity traffic environments. From a technological feasibility perspective, the use of a silicon substrate, dielectric patterning for strain engineering and high thermal budget tolerance enables facile integration of the phototransistor architecture with Si CMOS and MEMS platforms

## 4 Experimental Methods

**Device Fabrication:** To create patterned and non-patterned hBN/WS$_2$ devices on a p++ Si/SiO$_2$ substrate, a local bottom gate pad of ~35 μm × 40 μm was first fabricated using e-beam lithography (EBL), followed by metal deposition and lift-off. For this, the substrate was spin-coated with EL9 and PMMA A4 (e-beam resists) and annealed at 180 °C for 3 mins to remove solvents. The resist was exposed to an electron beam using the RAITH 150 Two electron beam lithography system and developed using a 1:1 mixture of 4-methyl-2-pentanone (MIBK) and 2-propanol (IPA). A 5 nm thick adhesive layer of titanium (Ti), followed by a 20 nm thick layer of gold (Au), were then deposited (to create the bottom electrode) using an AJA ATC sputter system. The lift-off process was carried out by submerging the sample in acetone and IPA.

An hBN flake was exfoliated using 3M Scotch tape and transferred onto a polydimethylsiloxane (PDMS) stamp attached to a glass substrate. The glass slide was connected to a micromanipulator under an Olympus BX-63 microscope to position and transfer the hBN flake onto the pre-fabricated Au bottom gate electrode. 450 nm diameter circular patterns of spin-coated e-beam resist (EL9/PMMA) (step 1, Figure S1) were then exposed using an e-beam, followed by development (step 2, Figure S1). Next, hBN was partially etched to a depth of approximately 15 nm (step 3, Figure S1) using a CHF$_3$+O$_2$ plasma under 0.2 bar base pressure for 3 minutes. Acetone was used to remove the e-beam resist from the unetched regions (step 4, Figure S1) to complete the fabrication of a nanopatterned (450 nm diameter, 15 nm deep) hBN gate dielectric.

Finally, a 1.8 nm-thick WS$_2$ flake was exfoliated and transferred onto the nanopatterned hBN layer (step 5, S1), followed by source/drain e-beam patterning and Ti/Au (5 nm/100 nm) sputter deposition and liftoff. Two transistors (T$_P$, T$_{NP}$) of 4×4 μm$^2$ area were fabricated such that they share a common Ti/Au source (S) and hBN/Au gate (G) but separate Ti/Au drain (D1, D2) electrodes.

**Device Characterization:** The device image in Figure 1(b) was captured using an Olympus BX-63 microscope. PL and Raman spectroscopy were performed using a Horiba HR 800 system with a 532 nm laser. Source, drain, and gate contacts of the patterned and non-patterned transistors were wire-bonded using gold wire to multiple large Au pads of a printed circuit board using a wedge bond system for optical characterization. Optoelectronic characterization was performed in ambient conditions under an Olympus BX-63 microscope using laser illumination from an NKT SuperK SELECT source with a wavelength range of 450 to 900 nm. The laser beam was illuminated through a 50X objective lens of the BX-63 Olympus microscope, and the radius of the laser spot size was set to 4 μm, such that it is incident on patterned (T$_P$) and non-patterned (T$_{NP}$) devices selectively. Time-resolved measurements were performed using square pulse excitation (10 V peak-to-peak) from the laser power controller unit of an Agilent 33220A function generator.

**Acknowledgments**

The authors thank Abin Varghese for his help with the optoelectronic measuring setup. The authors acknowledge the Indian Institute of Technology Bombay Nanofabrication Facility (IITBNF) for the usage of its device fabrication and characterization facilities and the Department of Science and Technology, Government of India, for funding this work through the project DST/NM/TUE/QM-8/2019 (G)2. Poulomi thanks the Indian Institute of Technology Bombay for the institute's post-doctoral fellowship.

**Conflict of Interest**

The authors declare no conflict of interest.

**Data Availability** The data supporting this study's findings are included within the article [and its Supporting Information].

**References**

[1] C. Li, C. Guo, L. Han, J. Jiang, M. M. Cheng, J. Gu, C. C. Loy, *IEEE Trans Pattern Anal Mach Intell* **2022**, *44*, DOI 10.1109/TPAMI.2021.3126387.

[2] R. Baba, K. Ueda, M. Okabe, *Dentomaxillofacial Radiology* **2004**, *33*, DOI 10.1259/dmfr/87440549.

[3] H. Rebecq, R. Ranftl, V. Koltun, D. Scaramuzza, *IEEE Trans Pattern Anal Mach Intell* **2021**, *43*, DOI 10.1109/TPAMI.2019.2963386.

[4] S. Wan, S. Goudos, *Computer Networks* **2020**, *168*, DOI 10.1016/j.comnet.2019.107036.

[5] K. Thakar, B. Rajendran, S. Lodha, *NPJ 2D Mater Appl* **2023**, *7*, DOI 10.1038/s41699-023-00422-z.

[6] G. Lu, B. Fei, *J Biomed Opt* **2014**, *19*, DOI 10.1117/1.jbo.19.1.010901.

[7] S. Ghosh, A. Varghese, K. Thakar, S. Dhara, S. Lodha, *Nat Commun* **2021**, *12*, DOI 10.1038/s41467-021-23679-8.

[8] Y. X. Zhou, Y. T. Lin, S. M. Huang, G. T. Chen, S. W. Chen, H. S. Wu, I. C. Ni, W. P. Pan, M. L. Tsai, C. I. Wu, P. K. Yang, *Nano Energy* **2022**, *97*, DOI 10.1016/j.nanoen.2022.107172.

[9] C. S. Boland, Y. Sun, D. G. Papageorgiou, *Nano Lett* **2024**, DOI 10.1021/acs.nanolett.4c03321.

[10] D. Bin Kim, J. Y. Kim, J. Han, Y. S. Cho, *Nano Energy* **2024**, *125*, DOI 10.1016/j.nanoen.2024.109551.

[11]  Y. Zhao, S. Jiao, S. Yang, Y. Nie, Y. Han, J. He, C. Tao, D. Wang, S. Gao, J. Wang, L. Zhao, *Advanced Materials* **2024**, DOI 10.1002/adma.202414649.

[12]  W. Peng, X. Wang, R. Yu, Y. Dai, H. Zou, A. C. Wang, Y. He, Z. L. Wang, *Advanced Materials* **2017**, *29*, 1606698, DOI 10.1002/adma.201606698.

[13]  C. Fang, T. Li, Y. Shao, Y. Wang, H. Hu, X. Li, X. Zeng, Y. Liu, Y. Hao, G. Han, *IEEE Trans Electron Devices* **2025**, DOI 10.1109/TED.2025.3545002.

[14]  E. M. F. Vieira, J. P. B. Silva, K. Gwozdz, A. Kaim, N. M. Gomes, A. Chahboun, M. J. M. Gomes, J. H. Correia, *Small* **2023**, *19*, DOI 10.1002/smll.202300607.

[15]  M. Kumar, D. Kim, H. Seo, *Small Methods* **2024**, *8*, DOI 10.1002/smtd.202300425.

[16]  Y. Dai, X. Wang, W. Peng, C. Xu, C. Wu, K. Dong, R. Liu, Z. L. Wang, *Advanced Materials* **2018**, *30*, DOI 10.1002/adma.201705893.

[17]  K. J. Lee, J. W. Min, B. Turedi, A. Y. Alsalloum, J. H. Min, Y. J. Kim, Y. J. Yoo, S. Oh, N. Cho, R. C. Subedi, S. Mitra, S. E. Yoon, J. H. Kim, K. Park, T. H. Chung, S. H. Jung, J. H. Baek, Y. M. Song, I. S. Roqan, T. K. Ng, B. S. Ooi, O. M. Bakr, *ACS Energy Lett* **2020**, *5*, DOI 10.1021/acsenergylett.0c01621.

[18]  V. Bhatt, M. Kumar, H. N. Kim, D. Yoo, J. H. Yun, M. J. Choi, *Nano Energy* **2025**, *133*, DOI 10.1016/j.nanoen.2024.110465.

[19]  J. Ma, M. Chen, S. Qiao, S. Guo, J. Chang, G. Fu, S. Wang, *Nano Energy* **2023**, *107*, DOI 10.1016/j.nanoen.2023.108167.

[20]  D. Wang, S. Ling, P. Hou, *ACS Appl Mater Interfaces* **2024**, *16*, 48576, DOI 10.1021/acsami.4c10005.

[21]  S. Sahoo, A. Varghese, A. Sadashiva, M. Goyal, J. Sakhuja, D. Bhowmik, S. Lodha, *ACS Nano* **2025**, DOI 10.1021/acsnano.5c00683.

[22]  Y. Zhang, H. Su, H. Li, Z. Xie, Y. Zhang, Y. Zhou, L. Yang, H. Lu, G. Yuan, H. Zheng, *Nano Energy* **2021**, *85*, DOI 10.1016/j.nanoen.2021.105968.

[23]  Y. Zhang, J. Chen, L. Zhu, Z. L. Wang, *Nano Lett* **2021**, *21*, 8808 DOI 10.1021/acs.nanolett.1c03171.

[24]  S. Guo, G. Zhang, Q. Wang, T. Zhang, S. Wang, L. Li, S. Qiao, *Laser Photon Rev* **2024**, *18*, DOI 10.1002/lpor.202400028.

[25]  L. Guo, X. Liu, L. Gao, X. Wang, L. Zhao, W. Zhang, S. Wang, C. Pan, Z. Yang, *ACS Nano* **2022**, *16*, DOI 10.1021/acsnano.1c09119.

[26]  S. Ghosh, A. Varghese, H. Jawa, Y. Yin, N. V. Medhekar, S. Lodha, *ACS Nano* **2022**, *16*, 4578 DOI 10.1021/acsnano.1c11110.

[27]  Q. Ma, G. Ren, K. Xu, J. Z. Ou, *Adv Opt Mater* **2021**, *9*, DOI 10.1002/adom.202001313.

[28]  K. Wang, B. Zhang, C. Yan, L. Du, S. Wang, *Nat Commun* **2024**, *15*, 9036 DOI 10.1038/s41467-024-53425-9.

[29]  A. Rani, S. D. Bu, *Current Applied Physics* **2024**, *66*, 1 DOI 10.1016/j.cap.2024.06.009.

[30]  Z. Wu, S. Xu, H. Lu, A. Khamoshi, G. Bin Liu, T. Han, Y. Wu, J. Lin, G. Long, Y. He, Y. Cai, Y. Yao, F. Zhang, N. Wang, *Nat Commun* **2016**, *7*, DOI 10.1038/ncomms12955.

[31] W. Huang, C. De-Eknamkul, Y. Ren, E. Cubukcu, *Opt Express* **2024**, *32*, 6076 DOI 10.1364/OE.510027.

[32] C. Torres-Torres, N. Perea-López, A. L. Elías, H. Rgutiérrez, D. Acullen, A. Berkdemir, F. López-Urías, H. Terrones, M. Terrones, *2d Mater* **2016**, *3*, DOI 10.1088/2053-1583/3/2/021005.

[33] E. Luppi, E. Degoli, M. Bertocchi, S. Ossicini, V. Véniard, *Phys Rev B Condens Matter Mater Phys* **2015**, *92*, DOI 10.1103/PhysRevB.92.075204.

[34] A. R. Khan, B. Liu, T. Lu, L. Zhang, A. Sharma, Y. Zhu, W. Ma, Y. Lu, *ACS Nano* **2020**, *14*, 15806 DOI 10.1021/acsnano.0c06901.

[35] W. Wu, L. Wang, Y. Li, F. Zhang, L. Lin, S. Niu, D. Chenet, X. Zhang, Y. Hao, T. F. Heinz, J. Hone, Z. L. Wang, *Nature* **2014**, *514*, 470 DOI 10.1038/nature13792.

[36] Z. Li, Y. Lv, L. Ren, J. Li, L. Kong, Y. Zeng, Q. Tao, R. Wu, H. Ma, B. Zhao, D. Wang, W. Dang, K. Chen, L. Liao, X. Duan, X. Duan, Y. Liu, *Nat Commun* **2020**, *11*, DOI 10.1038/s41467-020-15023-3.

[37] Q. Zhang, Z. Chang, G. Xu, Z. Wang, Y. Zhang, Z. Q. Xu, S. Chen, Q. Bao, J. Z. Liu, Y. W. Mai, W. Duan, M. S. Fuhrer, C. Zheng, *Adv Funct Mater* **2016**, *26*, DOI 10.1002/adfm.201603064.

[38] T. G. A. Verhagen, K. Drogowska, M. Kalbac, J. Vejpravova, *Phys Rev B Condens Matter Mater Phys* **2015**, *92*, DOI 10.1103/PhysRevB.92.125437.

[39] W. Wang, L. Zhou, S. Hu, K. S. Novoselov, Y. Cao, *Adv Funct Mater* **2021**, *31*, DOI 10.1002/adfm.202005053.

[40] P. Chaudhary, H. Lu, M. Loes, A. Lipatov, A. Sinitskii, A. Gruverman, *Nano Lett* **2022**, *22*, DOI 10.1021/acs.nanolett.1c04019.

[41] Z. D. Zhang, C. Cheng, S. Y. Yu, M. H. Lu, Y. F. Chen, *Phys Rev Appl* **2021**, *15*, DOI 10.1103/PhysRevApplied.15.034015.

[42] P. Chakrabarty, S. Sen, S. Sahoo, S. Lodha, in *IEEE Electron Devices Technology and Manufacturing Conference: Strengthening the Globalization in Semiconductors, EDTM 2024*, Institute Of Electrical And Electronics Engineers Inc., 2024, DOI 10.1109/EDTM58488.2024.10512332.

[43] C. Lee, B. G. Jeong, S. J. Yun, Y. H. Lee, S. M. Lee, M. S. Jeong, *ACS Nano* **2018**, *12*, DOI 10.1021/acsnano.8b04265.

[44] E. Stellino, B. D'Alò, E. Blundo, P. Postorino, A. Polimeni, *Nano Lett* **2024**, *24*, 3945, DOI 10.1021/acs.nanolett.4c00157.

[45] S. Han, J. Liu, A. I. Pérez-Jiménez, Z. Lei, P. Yan, Y. Zhang, X. Guo, R. Bai, S. Hu, X. Wu, D. W. Zhang, Q. Sun, D. Akinwande, E. T. Yu, L. Ji, *ACS Applied Materials and Interfaces* **2024**, *16*, 36735 DOI 10.1021/acsami.4c00092.

[46] A. R. Khan, T. Lu, W. Ma, Y. Lu, Y. Liu, *Adv Electron Mater* **2020**, *6*, DOI 10.1002/aelm.201901381.

[47] M. G. Bianchi, F. Risplendi, M. Re Fiorentin, G. Cicero, *Advanced Science* **2024**, *11*, DOI 10.1002/advs.202305162.

[48] A. K. Katiyar, Y. Kim, B. J. Kim, J. Choi, A. T. Hoang, J. D. Lee, J. H. Ahn, *Small* **2025**, DOI 10.1002/smll.202411378.

[49] D. Negi, M. Baishya, A. R. Moghe, S. Paul, S. Badola, S. Saha, *Small* **2025**, *21*, DOI 10.1002/smll.202412832.

[50] A. Varghese, A. H. Pandey, P. Sharma, Y. Yin, N. V. Medhekar, S. Lodha, *Nano Lett* **2024**, *24*, 8472, DOI 10.1021/acs.nanolett.4c00357.

[51] J. Zhang, Y. Li, X. Li, Y. Zhai, Q. Zhang, D. Ma, S. Mao, Q. Deng, Z. Li, X. Li, X. Wang, Y. Liu, Z. Zhang, X. Han, *Nat Commun* **2021**, *12*, DOI 10.1038/s41467-021-22447-y.

[52] M. Patel, H. H. Park, P. Bhatnagar, N. Kumar, J. Lee, J. Kim, *Nat Commun* **2024**, *15*, DOI 10.1038/s41467-024-47483-2.

[53] X. Zheng, M. Dong, Q. Li, Y. Liu, X. Di, X. Lu, J. Meng, Z. Li, *Adv Opt Mater* **2024**, *12*, DOI 10.1002/adom.202303177.

[54] Z. Wang, R. Yu, X. Wang, W. Wu, Z. L. Wang, *Advanced Materials* **2016**, *28*, DOI 10.1002/adma.201600884.

[55] Q. Guan, Z. K. Zhu, H. Ye, C. Zhang, H. Li, C. Ji, X. Liu, J. Luo, *Advanced Science* **2024**, DOI 10.1002/advs.202404403.

[56] S. Podder, J. Bora, K. B. Singh, D. Gogoi, B. Basumatary, A. R. Pal, *J Mater Chem A Mater* **2025**, DOI 10.1039/d5ta00063g.

[57] X. Yang, B. Zhou, M. Guo, Y. Liu, R. Cong, L. Li, W. Wu, S. Wang, L. Guo, C. Pan, Z. Yang, *Advanced Science* **2025**, DOI 10.1002/advs.202414422.

[58] K. Li, X. Wang, Y. Wu, W. Deng, J. Li, J. Li, Y. Zhao, Z. Chen, D. Yang, S. Yu, Y. Zhang, *InfoMat* **2025**, DOI 10.1002/inf2.70007.

[59] N. E. Silva, A. R. Jayakrishnan, A. Kaim, K. Gwozdz, L. Domingues, J. S. Kim, M. C. Istrate, C. Ghica, M. Pereira, L. Marques, M. J. M. Gomes, R. L. Z. Hoye, J. L. MacManus-Driscoll, J. P. B. Silva, *Adv Funct Mater* **2024**, DOI 10.1002/adfm.202416979.

[60] M. Kumar, H. Park, H. Seo, *Advanced Materials* **2024**, DOI 10.1002/adma.202406607.

[61] J. Zhu, Q. Cai, P. Shao, S. Zhang, H. You, H. Guo, J. Wang, J. Xue, B. Liu, H. Lu, Y. Zheng, R. Zhang, D. Chen, *Nat Commun* **2025**, *16*, 1186, DOI 10.1038/s41467-025-56617-z.

[62] D. Fu, Y. Ma, S. Wu, L. Pan, Q. Wang, R. Zhao, X. M. Zhang, J. Luo, *InfoMat* **2024**, DOI 10.1002/inf2.12602.

[63] Y. Wang, H. Ye, P. Wang, Z. Wu, Q. Guan, C. Zhang, H. Li, S. Chen, J. Luo, *Advanced Materials* **2024**, DOI 10.1002/adma.202409245.

[64] M. Kumar, M. Patel, J. Kim, D. Lim, *Nanoscale* **2017**, *9*, 19201, DOI 10.1039/C7NR07120E.

[65] J. Huang, Q. Li, X. Lu, J. Meng, Z. Li, *Adv Mater Interfaces* **2022**, *9*, DOI 10.1002/admi.202200327.

[66] R. Ahmadi, A. Abnavi, H. Ghanbari, H. Mohandes, M. R. Mohammadzadeh, T. De Silva, A. Hasani, M. Fawzy, F. Kabir, M. M. Adachi, *Nano Energy* **2022**, *98*, DOI 10.1016/j.nanoen.2022.107285.

[67] A. Abnavi, R. Ahmadi, H. Ghanbari, M. Fawzy, M. R. Mohammadzadeh, F. Kabir, M. M. Adachi, *Adv Opt Mater* **2024**, *12*, DOI 10.1002/adom.202302651.

[68] M. Kumar, A. Saravanan, S. C. Chen, B. R. Huang, H. Sun, *ACS Appl Mater Interfaces* **2025**, DOI 10.1021/acsami.5c02944.

# Supporting Information

## Electrostatically Controlled Pyrophototronic Effect Enabled Accident Alert System using a Strain-Polarized WS$_2$ Phototransistor

Poulomi Chakrabarty,[1] Sera Sen,[1] Anwesha Chakraborty,[2] Abhay Anand VS,[3] Srilagna Sahoo,[1] Anshuman Kumar,[3] Debjani Karmakar,[2,4,5] and Saurabh Lodha[1*]

[1]Department of Electrical Engineering, Indian Institute of Technology Bombay, India
, [2]Technical Physics Division, BARC, Mumbai 400085, India, [3]Department of Physics, Indian Institute of Technology Bombay, India, Homi Bhabha National Institute, [4]Anushaktinagar, Mumbai, 400094, India, Department of Physics and Astronomy, [5]Uppsala University, Box 516, SE-75120, Uppsala, Sweden
Email: slodha@ee.iitb.ac.in

### S1 Device Fabrication Process Flow

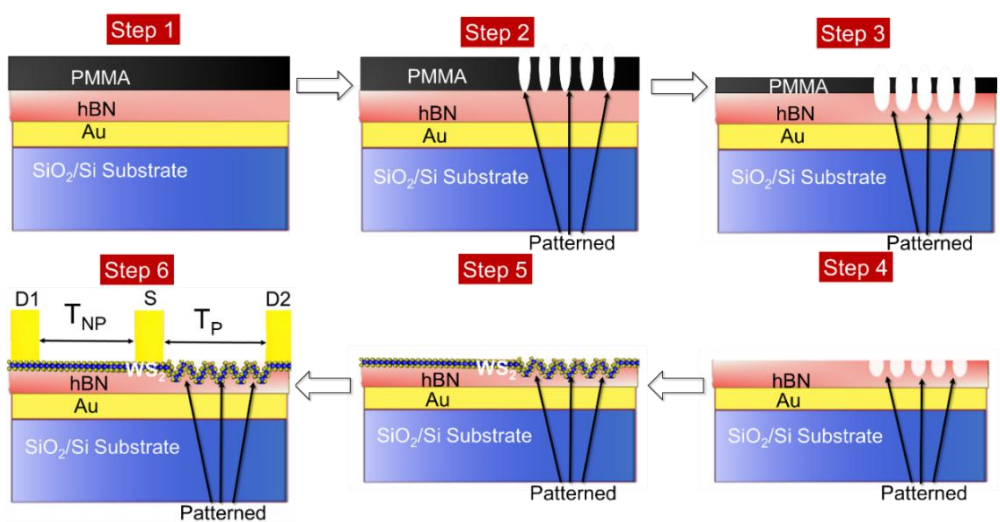

*Figure S1. Device fabrication process flow shown through cross-section schematics.*

### S2 Raman and Photoluminescence (PL) Spectra and Spatial Maps

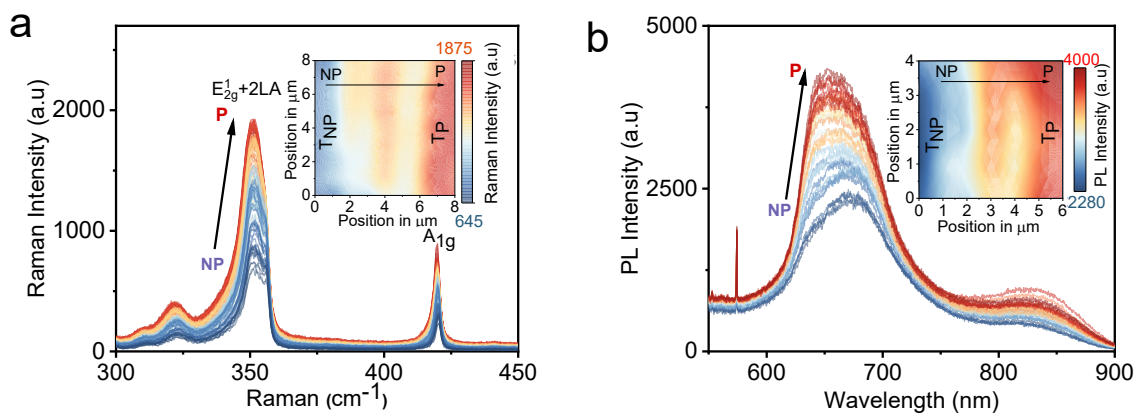

*Figure S2. (a) Raman and (b) PL mapping spectra of the devices $T_P$ and $T_{NP}$, respectively. The insets show the area maps corresponding to the spectra.*

## S3 FDTD Calculations of the Local Electric Field

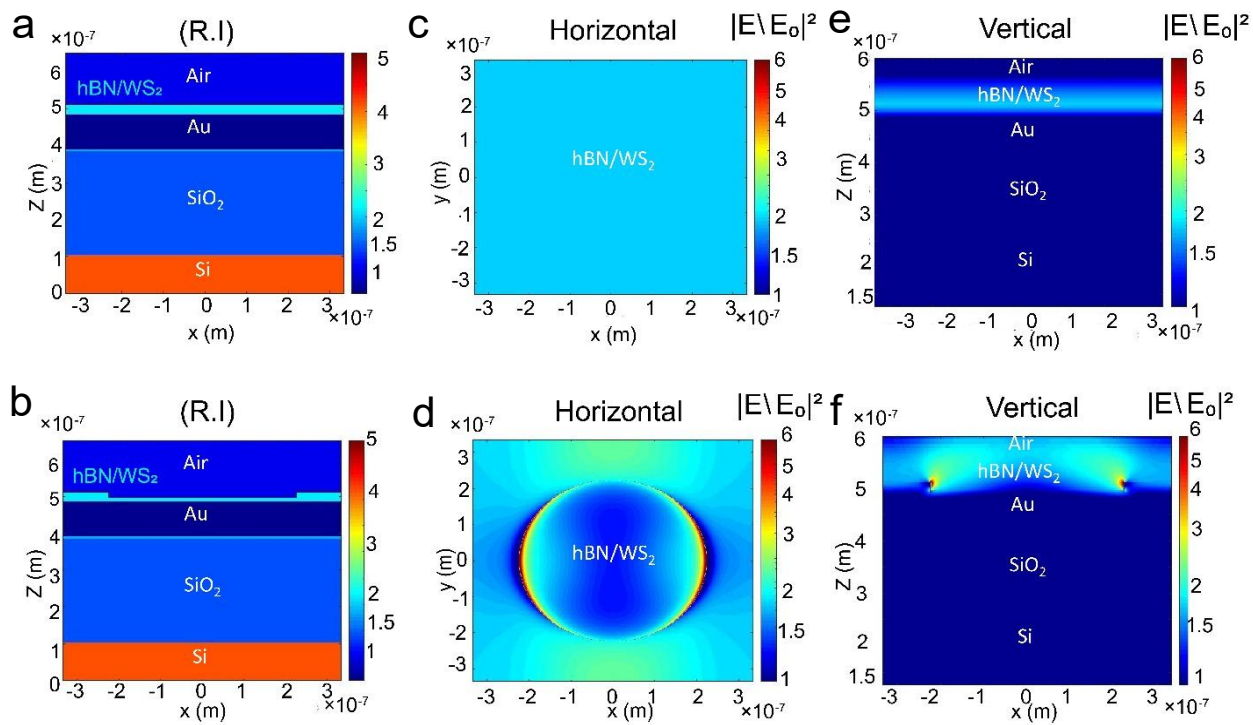

*Figure S3. Refractive indices of various materials used in the (a) non-patterned device: $T_{NP}$, and (b) the patterned device: $T_P$. Horizontal electric field $|E/E_0|^2$ for the (c) non-patterned device: $T_{NP}$, and the (d) patterned device: $T_P$. Vertical electric field $|E/E_0|^2$ for the (e) non-patterned device: $T_{NP}$, and the (f) patterned device: $T_P$.*

To provide more insight into enhanced PL emission from region P compared to region NP, we have estimated the PL enhancement factor (EF) of the strain localized excitons under 532 nm excitation using the formula EF = $|E/E_0|^2$ × (QY/QY$_0$), where $|E/E_0|^2$ is the electric field intensity enhancement at the excitation wavelength. QY/QY$_0$ is the change in the emission quantum yield, indicating the modulation of the spontaneous decay rate in the system. Region P and NP share the same flake, and hence, we can assume that the QY doesn't change substantially with hBN patterning. Based on finite difference time domain (FDTD) simulations, a high EF of ~6 is obtained at the contour of the WS$_2$ flakes placed on the nanopatterned hBN. Coupling of the incident light with the subwavelength patterns and strained WS$_2$ results in substantial enhancement of PL emission from the strain-localized excitons at room temperature.

# S4-S7 Density Functional Theory Calculations for Strained WS$_2$

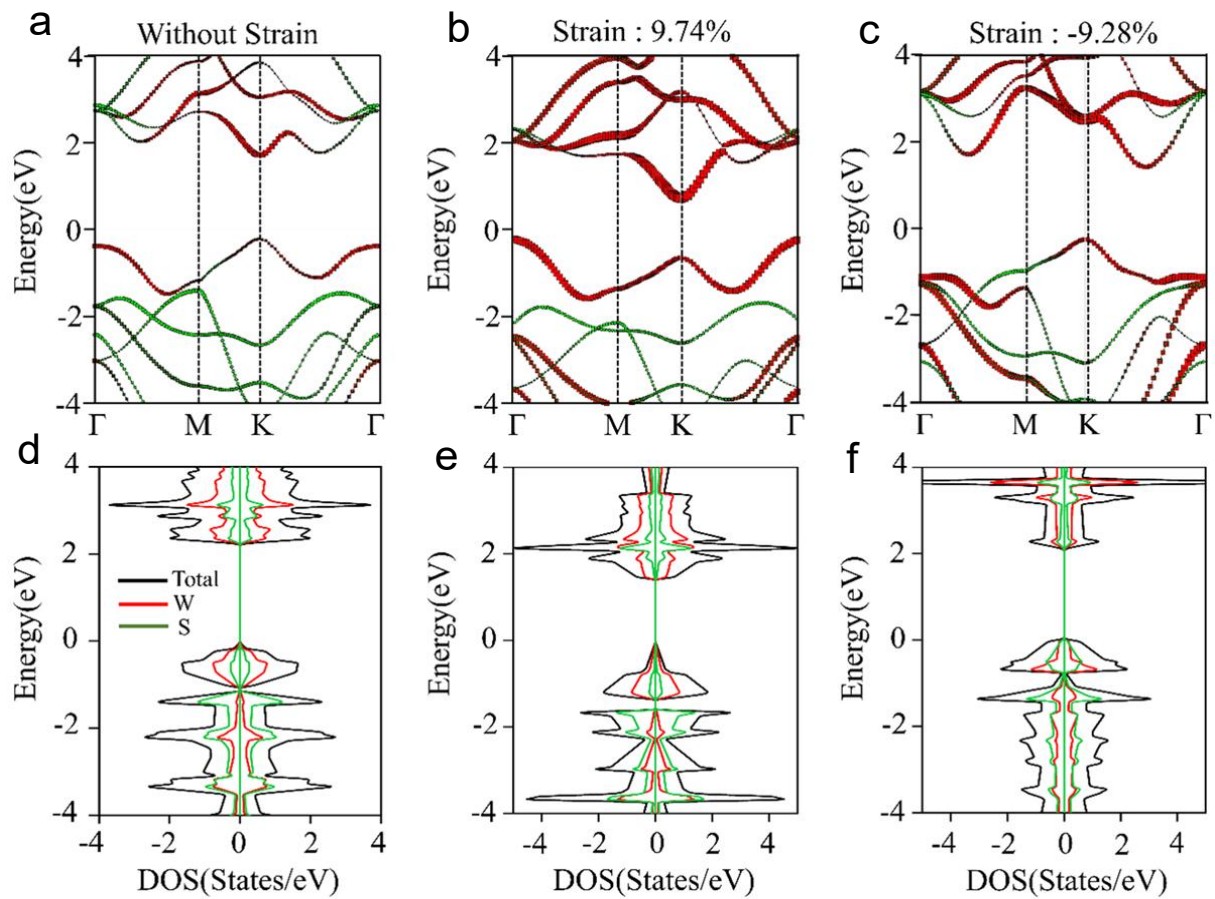

*Figure S4. Comparison of band structure of 1-layer (1L) WS$_2$ (a) without strain, (b) tensile strain, and (d) compressive strain and DOS of 1L WS$_2$ (d) without strain and with e) tensile and, (f) compressive strain.*

Figure S4 presents the band structure along with total and partial density of states (DOS) for both tensile and compressive strain in 1L WS$_2$. Under both types of strain, 1L WS$_2$ transitions from direct to indirect band gap. However, while tensile strain impacts both the Γ and K points changing the valence band maximum (VBM) and the conduction band minimum (CBM) to reduce the gap, compressive strain primarily impacts the CBM with an increase in band gap compared to the unstrained condition.[1] In addition, hybridization of W-5$d$ and S-3$p$ states doesn't change significantly with tensile strain, whereas compressive strain makes the S-3$p$ states more prominent at the band edges.

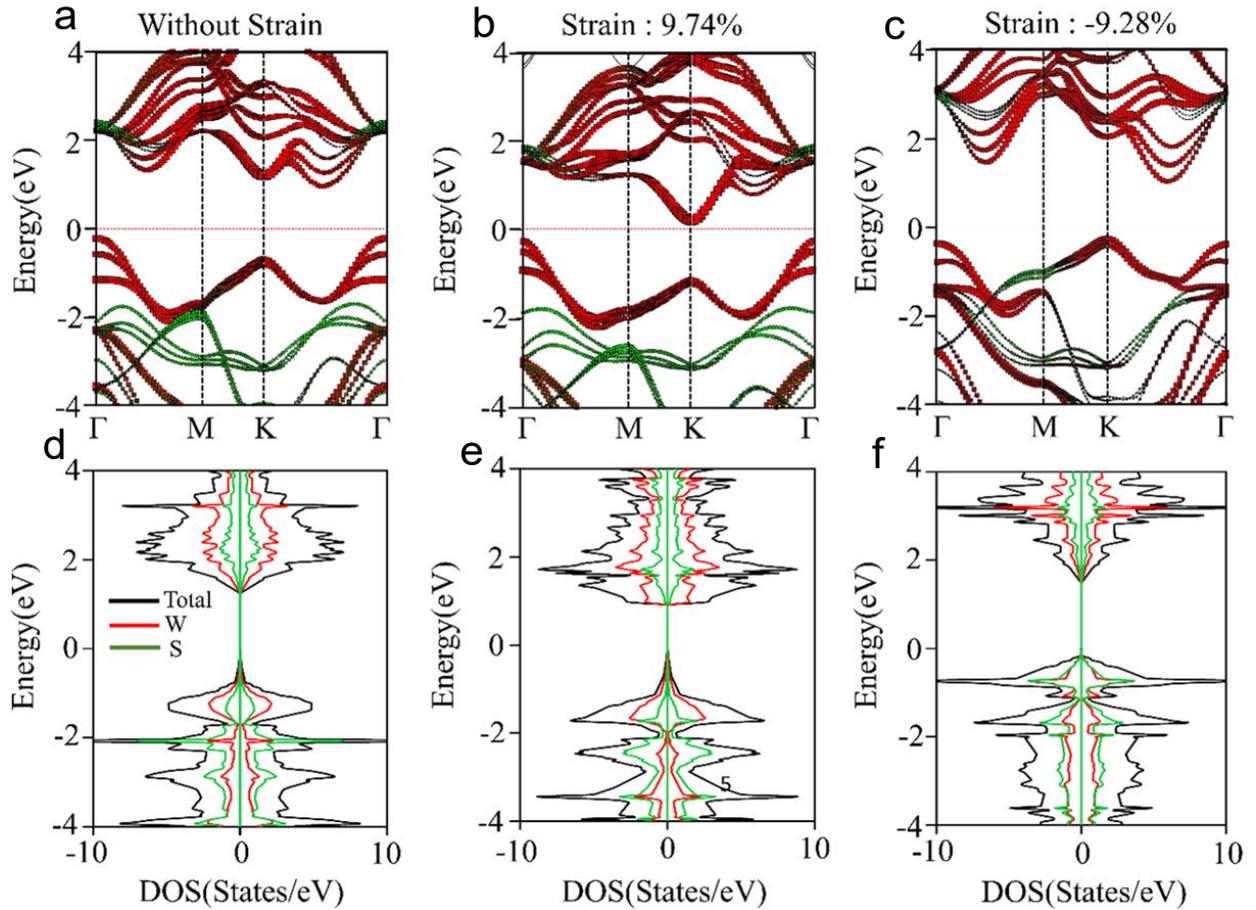

*Figure S5. Comparison of band structure of 3-layer (3L) WS$_2$ (a) without strain, (b) tensile strain, and (d) compressive strain and DOS of 3L WS$_2$ (d) without strain and with e) tensile and, (f) compressive strain.*

Figure S5 represents a similar comparison of the band structure and DOS for 3L-WS$_2$. In 3L, the direct band gap in the ambient condition reduces significantly for tensile strain with a decrease of *p-d* hybridization near CBM. Under compressive strain, there is an increase in both the band-gap and the hybridizations at the band edges.

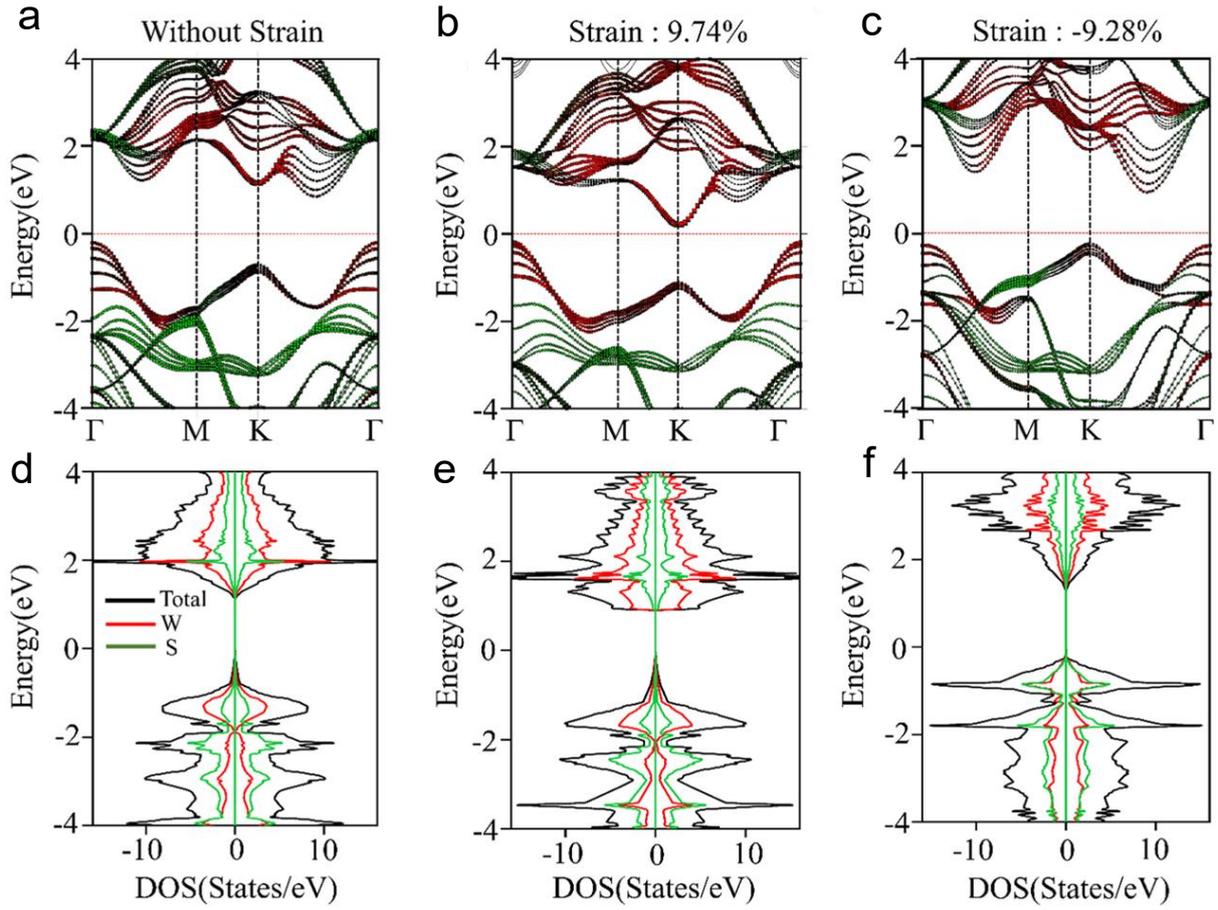

*Figure S6. Comparison of band structure of 5-layer (5L) WS$_2$ (a) without strain, (b) tensile strain, and (d) compressive strain and DOS of 5L WS$_2$ (d) without strain and with e) tensile and, (f) compressive strain.*

Figure S6 depicts a similar comparison of the band structure and DOS for 5L WS$_2$. In five layers, the band structure for ambient conditions and under tensile strain shows a similar trend as 1L and 3L, but with an increase in DOS. Under compressive strain, there is also an increase in both the band gap and the hybridizations at the band edges.

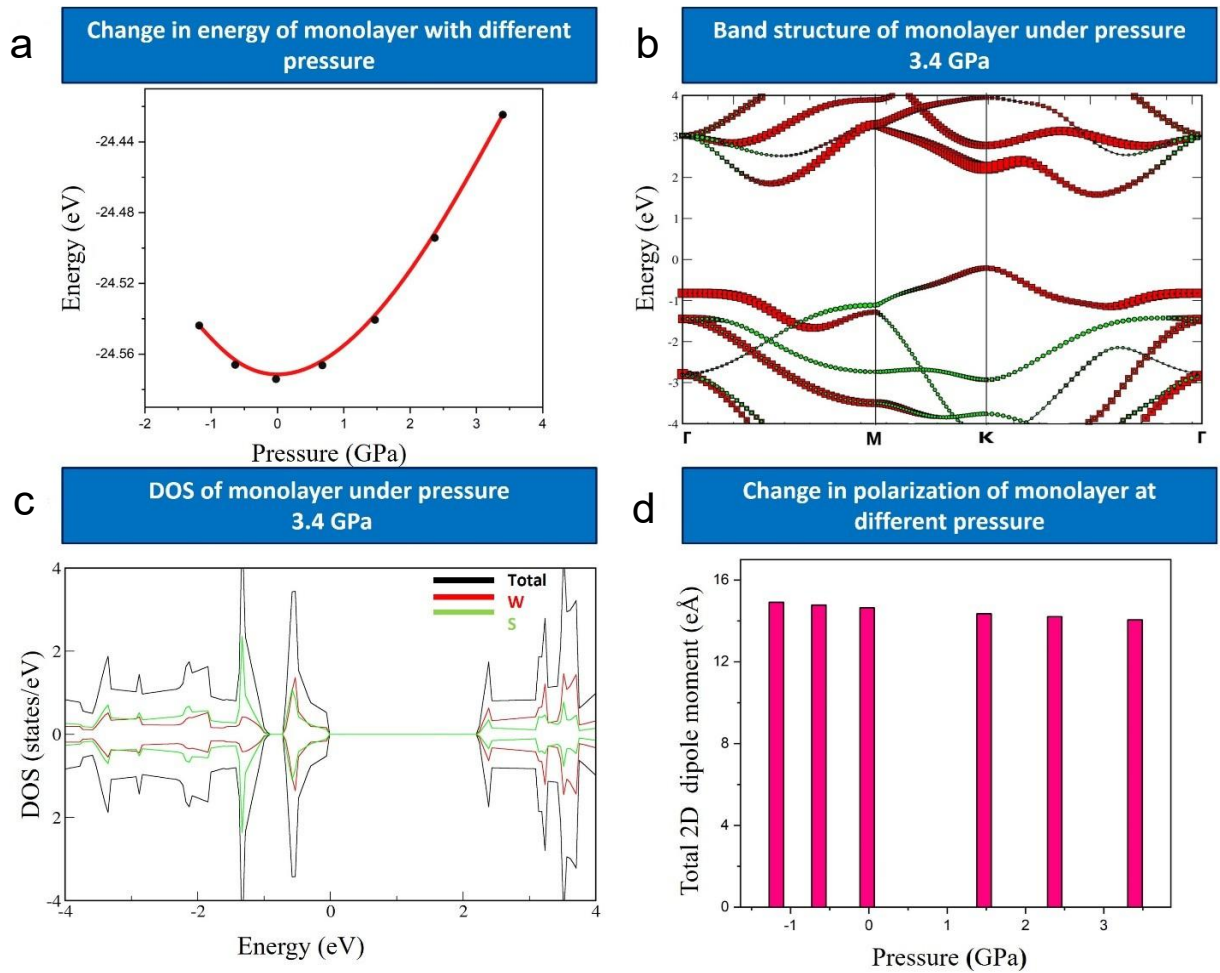

*Figure S7. (a) Energy versus pressure equation of state fitting under volumetric strain, (b) band structure, and (c) DOS of monolayer at a pressure of 3.4 GPa, (d) variation of electronic dipole moment at different pressures.*

In addition, for 1L, we have also investigated the impact of both compressive and expansive volumetric strain, and the results are presented in Figure S7. Under expansive strain, there is an increase in polarization. With compression of volume, the polarization decreases, and the system undergoes a direct-to-indirect crossover. Therefore, the theoretical investigation implies that structural deformations induced by both biaxial and volumetric strain may induce a charge imbalance within the few-layered $WS_2$ system and thus may result in a non-zero polarization leading to the piezoelectric properties within it.

## S8 Dark Output I-V Characteristics of $T_P$ and $T_{NP}$ Phototransistors

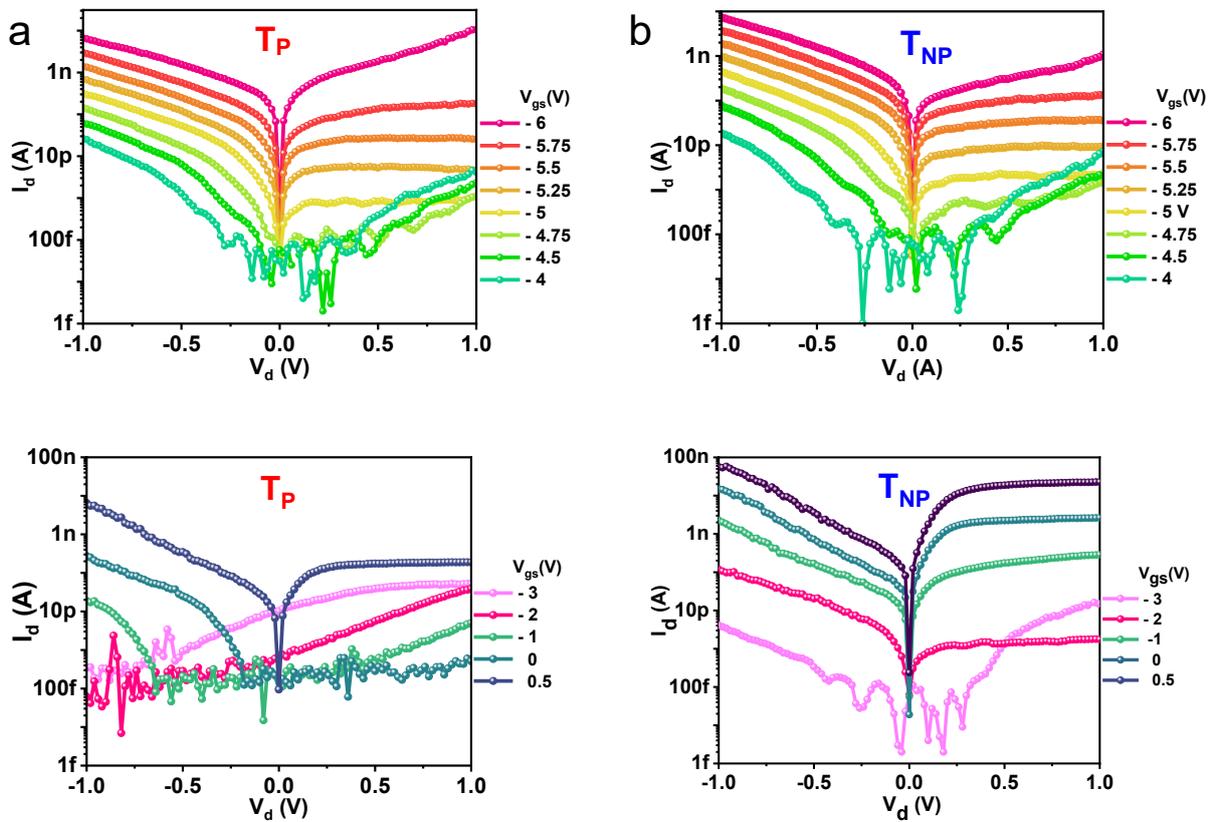

*Figure S8. Electrical characterization of patterned and non-patterned phototransistors. The output ($I_d$-$V_{ds}$) characteristics of transistors (a) $T_P$ (patterned) and (b) $T_{NP}$ (non-patterned) at $V_d = 1\ V$ under dark conditions for varying gate-to-source voltages ($V_{gs}$).*

Output characteristics ($I_d$ vs drain voltage $V_d$) in Figure S8 indicate Schottky barrier-dominated transport across the S/D contacts to the $WS_2$ channel.

## S9 Electrical Device Characterization: Dark and Steady-state Illumination

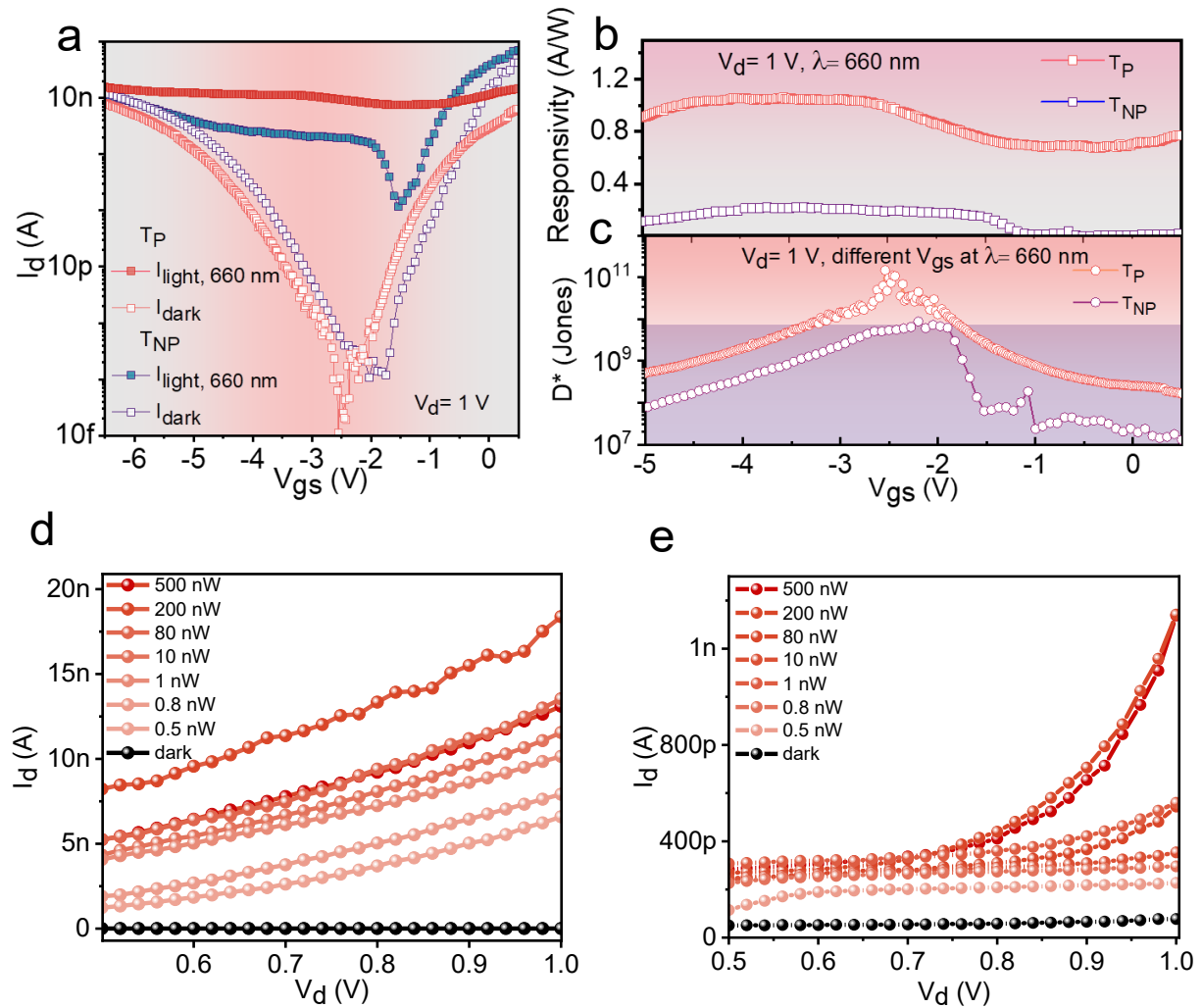

*Figure S9. Steady-state optoelectronic characterization of $T_P$ (patterned) and $T_{NP}$ (non-patterned) phototransistors:* (a) Drain current vs gate-to-source voltage ($I_d$ -$V_{gs}$) transfer characteristics of $T_P$ and $T_{NP}$ at $V_d$ = 1 V under dark and 660 nm light illumination showing ambipolar transport characteristics. Both n-branch and p-branch on-state currents in the dark condition are attributed to contact Schottky barrier tunnelling due to gate-induced barrier reduction and thinning. (b) Photoresponsivity and (c) detectivity of $T_P$ and $T_{NP}$ show gate-tunable photo response at $V_d$ = 1 V at 10 nW of 660 nm laser wavelength illumination. $I_{light}$ vs incident power data for (d) $T_P$ and (e) $T_{NP}$ at $V_{gs}$= -3V and $V_d$= 1 V at 660 nm wavelength of light.

Figure S9(a) shows the ambipolar transfer (drain current $I_d$ vs gate-to-source voltage $V_{gs}$) curves for both transistors under dark ($I_{dark}$) and light exposure ($I_{light}$), similar to previously reported ambipolar FETs.[2] The difference in dark currents of the two transistors can be attributed to an electrostatic threshold voltage ($V_{th}$) shift, likely due to positive fixed charge induced in the hBN dielectric during plasma etching. The $V_{th}$ is lower (higher) in the n-dominated (p-dominated) transport regime of $T_P$ compared to $T_{NP}$. The higher photo current $I_{ph}$ ($I_{ph}= I_{light}-I_{dark}$) for $T_P$ is due to enhanced light-matter interaction resulting from hBN nanopatterning-induced light trapping and enhanced optical path length, which increases light absorption and photocurrent. This can be correlated with enhanced PL intensity in $T_P$ compared to $T_{NP}$ (Figure 1(e)) and a value of 6 for the PL enhancement factor (EF) of $T_P$ with respect to $T_{NP}$, as obtained from FDTD simulations in Figure S3. Figure S9(b) depicts the photoresponsivity, $R(\lambda)=\frac{I_{ph}}{P_{opt}}$ where $P_{opt}$ is the optical power at wavelength $\lambda$, of $T_P$ and $T_{NP}$ for varying $V_{gs}$ at fixed $V_d= 1$ V. $T_P$ and $T_{NP}$ show a peak responsivity of 1.04 A/W and 0.20 A/W, respectively, for $V_{gs}= -3$ V at $P_{opt}$ of 10 nW. Peak $R(\lambda)$ of $T_P$ is thus ~5X higher than that of $T_{NP}$, which can be attributed to the primarily higher $I_{light}$. $R(\lambda)$ of $T_P$ is higher than for $T_{NP}$, even for the n-dominated branch, despite larger $I_{dark}$. This indicates that the patterning-induced enhanced light-matter interaction plays a more dominant role in the improved photodetection performance of $T_P$. It is also evident from Figure S9(a) that under light illumination, the light/dark current ratio (maximum $I_{light}/I_{dark}$ in $T_{NP}$ ~$10^4$ and ~$10^5$ in $T_P$ at $V_{gs}= -3$ V and $V_d= 1$ V) is tunable with applied $V_{gs}$. Gate tunability of $I_{light}/I_{dark}$ adds an important knob to control the photoresponse for detection of weak and strong light illumination. Shot noise-limited specific detectivity ($D^*(\lambda)=\frac{R(\lambda)\times\sqrt{A}}{\sqrt{2qI_{dark}}}$)[3], where q is the electron charge and A is the active area of the device, is plotted vs $V_{gs}$ in Figure S9(c). Peak specific detectivity is estimated to be $1.6\times10^{11}$ Jones and $8.7\times10^9$ Jones at $V_{gs}= -3$ V for $T_P$ and $T_{NP}$, respectively. The higher R for $T_P$ results in a larger $D^*$ compared to $T_{NP}$ for the entire range of $V_{gs}$. In summary, $T_P$ shows improved steady-state photodetection performance as compared to $T_{NP}$, primarily due to enhanced light-matter interaction. The steady-state $I_{light}$ vs incident power data for $T_P$ and $T_{NP}$ at $V_{gs}= -3$ V, $V_d = 1$ V, and $\lambda= 660$ nm are shown in Figure S9(d) and Figure S9(e), respectively.

## S10 Illumination Power-dependent Detectivities due to Pyrophototronic Effect

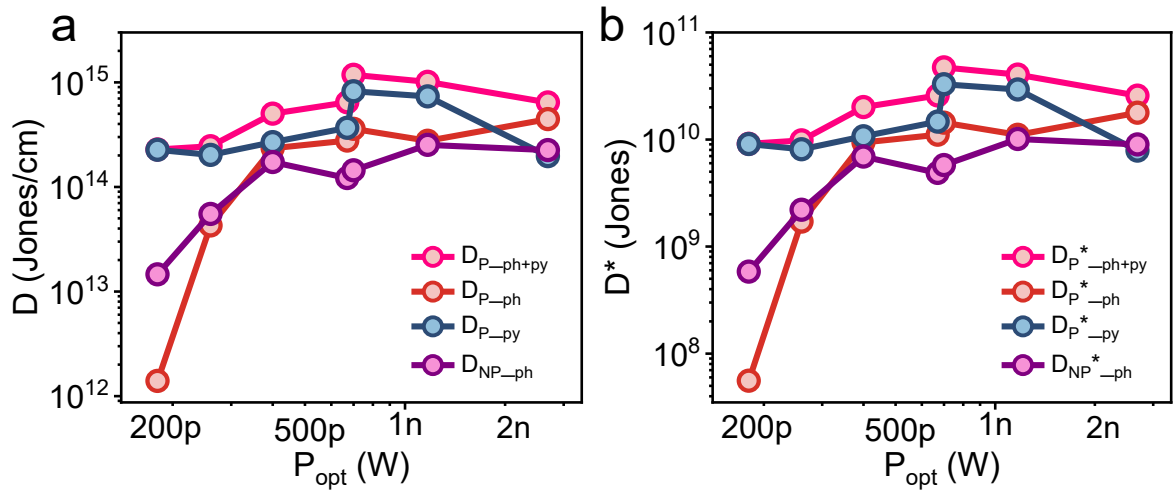

*Figure S10. Comparison of optical power-dependent (a) pyro+photodetectivity ($D_{p\_ph+py}$), photodetectivity ($D_{p\_ph}$), pyrodetectivity ($D_{p\_py}$) of $T_P$ with the photodetectivity ($D_{NP\_ph}$) of $T_{NP}$ and (b) specific pyro+photodetectivity ($D_p^*{}_{\_ph+py}$), specific photodetectivity ($D_p^*{}_{\_ph}$), specific pyrodetectivity ($D_p^*{}_{\_py}$) of $T_P$ with the specific photodetectivity ($D_{NP}^*{}_{\_ph}$) of $T_{NP}$.*

The detectivity and specific detectivity are calculated using equations ($D(\lambda) = \frac{R(\lambda)}{\sqrt{2qI_{dark}}}$) and ($D^*(\lambda) = \frac{R(\lambda) \times \sqrt{A}}{\sqrt{2qI_{dark}}}$).

## References

[1] A. Rawat, N. Jena, Dimple, A. De Sarkar, *J Mater Chem A Mater* **2018**, *6*, 8693, DOI 10.1039/C8TA01943F.

[2] Z. Wang, Q. Li, Y. Chen, B. Cui, Y. Li, F. Besenbacher, M. Dong, *NPG Asia Mater* **2018**, *10*, DOI 10.1038/s41427-018-0062-1.

[3] S. Ghosh, A. Varghese, K. Thakar, S. Dhara, S. Lodha, *Nat Commun* **2021**, *12*, DOI 10.1038/s41467-021-23679-8.